\documentclass[conference]{IEEEtran}
\IEEEoverridecommandlockouts

\usepackage{multirow}
\usepackage{xcolor}
\usepackage[framemethod=TikZ]{mdframed}
\usepackage{tcolorbox}
\usepackage{float}
\usepackage{xspace}
\newtcolorbox{mybox}{colback=red!5!white,colframe=red!75!black}
\usepackage{flushend}
\usepackage{amssymb} 
\usepackage{amsmath}
\usepackage{booktabs}
\usepackage{graphicx}   %
\usepackage[numbers]{natbib}
\usepackage{array}
\usepackage{url}
\usepackage[small,bf]{caption}
\usepackage[skip=0pt]{subcaption}
\usepackage[]{footmisc}

\mdfdefinestyle{MyFrame}{
    linecolor=black,
    outerlinewidth=0.5pt,
    roundcorner=5pt,
    innertopmargin=2pt,
    innerbottommargin=2pt,
    innerrightmargin=-10pt,
    innerleftmargin=-15pt,
    leftmargin = -10pt,
    rightmargin = -10pt,
    backgroundcolor=gray!10,
    userdefinedwidth=220pt
}

\AtBeginDocument{%
  }

\begin{document}

\date{}

\newcommand{\systemname}[0]{\textsc{Loki}\xspace}
\newcommand{\descr}[1]{\smallskip\noindent\textbf{#1}}

\title{\systemname: Proactively Discovering Online Scam \\Websites by Mining Toxic Search Queries\textsuperscript{*}\thanks{\textsuperscript{*}Published in the Proceedings of the 33rd Network and Distributed System Security Symposium (NDSS 2026) -- please cite the NDSS version.}}

 \author{\IEEEauthorblockN{Pujan Paudel}
  \IEEEauthorblockA{Boston University\\
  ppaudel@bu.edu}
  \and
  \IEEEauthorblockN{Gianluca Stringhini}
  \IEEEauthorblockA{Boston University\\
  gian@bu.edu}}

\maketitle

\begin{abstract}
Online e-commerce scams, ranging from shopping scams to pet scams, globally cause millions of dollars in financial damage every year.
In response, the security community has developed highly accurate detection systems able to determine if a website is fraudulent.
However, finding candidate scam websites that can be passed as input to these downstream detection systems is challenging: relying on user reports is inherently reactive and slow, and proactive systems issuing search engine queries to return candidate websites suffer from low coverage and do not generalize to new scam types.
In this paper, we present \systemname, a system designed to identify search engine queries likely to return a high fraction of fraudulent websites.
\systemname implements a keyword scoring model grounded in Learning Under Privileged Information (LUPI) and feature distillation from Search Engine Result Pages (SERPs).
We rigorously validate \systemname across 10 major scam categories and demonstrate a 20.58 times improvement in discovery over both heuristic and data-driven baselines across all categories.
Leveraging a small seed set of only 1,663 known scam sites, we use the keywords identified by our method to discover 52,493 previously unreported scams in the wild.
Finally, we show that \systemname generalizes to previously-unseen scam categories, highlighting its utility in surfacing emerging threats. %

\end{abstract}

\section{Introduction}
Online scam websites, where miscreants set up legitimate-looking sites designed to directly defraud users, are on the rise~\cite{bitaab2023beyond}.
Regulatory bodies like the Australian Competition and Consumer Commission and the UK Financial Conduct Authority report a rising trend in this threat~\cite{muzammil2025poorest}, and the US Federal Trade Commission (FTC) recently reported that fraudulent shopping scams were the second most reported fraud type in the country in 2024, accounting for over \$432 million in consumer losses~\cite{ftc2024csn}.
As more types of transactions switch to the online domain, new fraud vectors emerge, including pet scams, fake charities, and investment fraud.
Therefore, identifying scam websites in a quick and proactive way is of paramount importance to keep users safe.

The problem of detecting scam websites has attracted significant interest from the security community, which has developed several systems to determine if a website is fraudulent or not~\cite{bitaab2023beyond,kotzias2023scamdog,nakano25dimva}. 
These systems showed that developing an oracle able to accurately classify scam websites is possible.
One of the main challenges, paradoxically, is to find websites that are candidate to being fraudulent, which will then be provided as input to said oracle.
The most straightforward avenue to find potential scam websites is by issuing search engine queries, where we aim to identify queries that maximize the number of scam websites returned.
To measure the effectiveness of search queries in surfacing fraudulent websites, we introduce the concept of \emph{toxicity} and \emph{expansion}, which we define as the empirical propensity of a search query to surface scam websites within its corresponding search engine result pages (SERPs).
For example, consider two queries related to the category of cryptocurrency, ``Double my bitcoin quickly'' ($Q_a$), and ``safe ways to buy bitcoins'' ($Q_b$).
When issued to a commercial search engine such as Bing, $Q_a$ yields 6 out of 20 results that correspond to scam websites, while $Q_b$ yields only 2 out of 20 scam websites, with the majority of links pointing to reputable or authoritative sources.
In other words, $Q_a$ exhibits higher toxicity ($0.3$) than $Q_b$ ($0.1$).
Intuitively, $Q_a$ allows to discover more scam websites (expansion of $6$) than $Q_b$ (expansion of $2$).

Previous works issue queries based on manually curated or heuristic-guided domain-specific lists of keywords, for example, to identify cryptocurrency~\cite{li2023double} and investment scams~\cite{muzammil2025poorest}.
Another line of work uses static methods like TF-IDF and topic modeling approaches~\cite{blei2003latent} over the HTML of previously identified scam websites to extract search queries and discover a broader set of scam websites~\cite{liu2023understanding,srinivasan2018exposing}.
While these techniques offer scalability and pathways for automation, they suffer from critical limitations.
In particular, these methods exhibit a strong bias toward capturing brand, business, or entity-specific lexical cues that are closely tied to the seed dataset, rather than identifying more general-purpose linguistic indicators reflective of the underlying \textit{modus operandi} of scam campaigns.
This limits the generalizability of the extracted keywords across scam types and reduces their capacity to support broad discovery, limiting their toxicity and subsequent expansion.%

An alternative approach is to issue search queries that latently encode operational patterns of scam websites (i.e., urgency, scarcity, or high-return incentives), or artifacts introduced by black hat search engine optimization (SEO) techniques~\cite{lu2011surf}.
While this assumption may hold within certain domains and specific types of scams, especially those characterized by aggressive SEO abuse and unintended brand association (e.g., typosquatting attacks), it does not generalize across the broader landscape of online scams.%
Another common approach in the literature relies on reactive data sources and online communities to probe and classify scam websites, for example, victim-reported complaints, subreddit submissions (e.g., \textit{/r/scams}), and threat intelligence feeds from commercial platforms (e.g., Palo Alto Networks, Norton Security, etc.)
While such sources are often grounded in real-world harm and offer highly precise labels, they suffer from delayed reporting, limited coverage, and implicit biases towards the types of scams that only end up being reported in online communities.
Most importantly, these approaches are by design retrospective, as they detect scams only after users have encountered and reported them.

These limitations highlight the necessity for query mining systems that can identify queries agnostic to content promotion mechanisms and actually capture scam-related cues that can generalize across diverse scam categories.
In this paper, we overcome the limitations of static and heuristic-based query extraction systems by developing \systemname, a system that enables the systematic querying of search engines to proactively identify scam websites. 
\systemname is designed to maximize the toxicity of search queries, and uses this metric to learn how to rank queries and issue search queries that are likely to return the most scam websites, to be fed to a high precision scam detection oracle.
\systemname can score candidate query keywords across multiple business categories for toxicity scores.
These predicted keyword scores for each category can then be ranked and prioritized for querying SERP, enabling proactive and efficient discovery of scam websites.
We rigorously evaluate \systemname in its ability to systematically identify the most toxic keywords to discover scam websites, and verify its generalization capabilities across 10 different categories.
To this end, we make the following key contributions:
\begin{itemize}
\item \textbf{Empirical analysis of heuristic-based sampling strategies:} We conduct a systematic comparison of query toxicity and expansion across existing attribute-based and entity-based query sampling methods.
Our analysis reveals that these approaches are brittle, exhibit limited generalization across scam categories, and consistently underperform in identifying high-toxicity queries.

\item \textbf{Data-Driven Keyword Scoring Framework:} We formulate a novel data-driven approach for query toxicity prediction.
By leveraging Learning Under Privileged Information (LUPI) and feature distillation,
\systemname learns to approximate the toxicity of search queries using privileged features derived from SERPs.
This enables scalable and guided discovery of scam domains by scoring and ranking large pools of candidate queries without requiring SERP access at inference time.

\item \textbf{Cross-Domain Validation Across Major Scam Categories:} We validate our approach across high-impact categories of online e-commerce scams known to cause significant real-world financial harm.
Our approach consistently outperforms both heuristic and data-driven keyword sampling baselines in identifying high-toxicity queries, demonstrating strong generalization and robustness across scam domains.

\item \textbf{Scalable Scam Discovery at Internet Scale:} Starting from a small seed set of approximately 1,500 known scam domains, \systemname discovers over 52,493 new scam domains, demonstrating its efficacy, scalability, and practical utility in real-world proactive scam detection.

\end{itemize}

\begin{figure*}{}{}
\scalebox{0.9}{
  \centering
    \includegraphics[width=\linewidth]{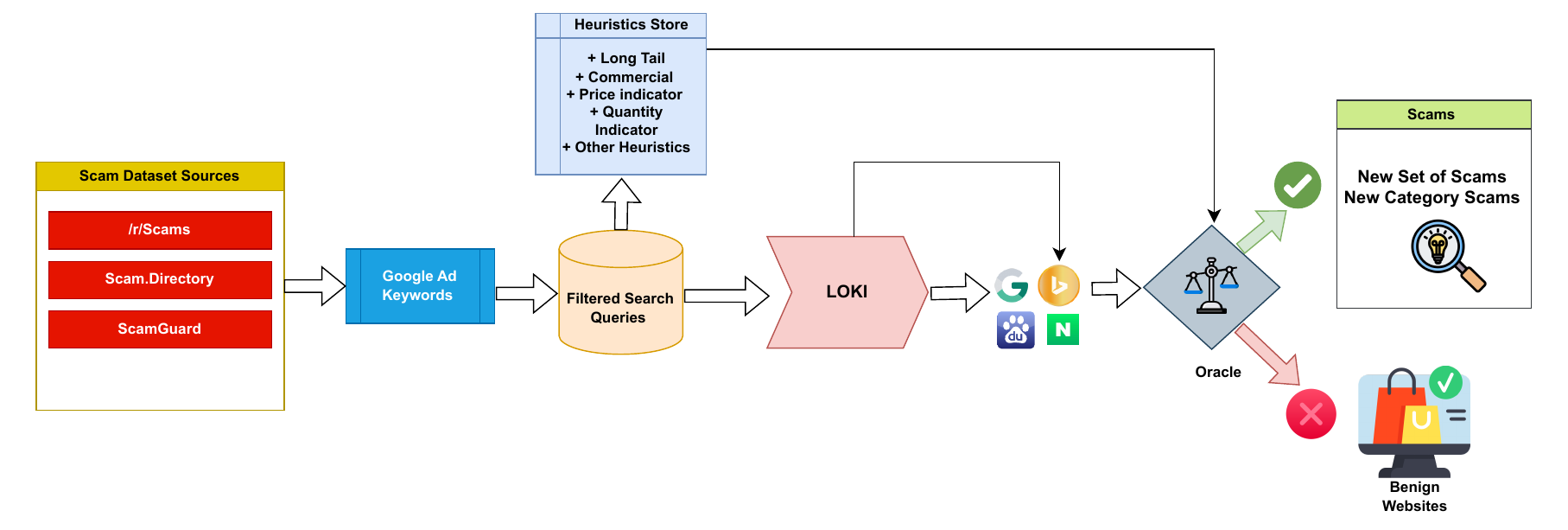}}
    \caption{Overview of our system pipeline.}
  \label{fig:system_pipeline}
\end{figure*}

Techniques such as LUPI and feature distillation are established frameworks in the field of Machine Learning (ML).
However, these techniques have not been previously applied in the context of ML for security. 
To the best of our knowledge, our work provides the first application using SERPs for both of these frameworks, including a novel application of these frameworks for identifying scam websites.
A visual summary of the contributions of our work is presented in Figure~\ref{fig:system_pipeline}.

\section{Datasets}
To build and evaluate our system, we curate four datasets.
First, we build ground truth datasets of scam and benign websites, which we will use to build an oracle to enable the downstream evaluation of \systemname's performance (see Section~\ref{sec:oracle_model}).
Next, we curate a dataset of Google Ad Keyword suggestions and SERP results, which we will use to validate our methodology and compare various baselines (see Section~\ref{sec:motivation}) and to build our \systemname system (see Section~\ref{sec:system}).
To encourage reproducibility and allow researchers to build on our work, we publicly release our datasets and models\footnote{https://doi.org/10.5281/zenodo.17049965}.
In this section, we describe these datasets in detail.

\subsection{Scam websites}
\label{sec:dataset_scam}
Collecting an up-to-date and comprehensive set of scam websites aggregated from multiple sources serves two primary purposes: i)~it enables the construction of a diverse training set for our oracle classifier by capturing a broad spectrum of scam categories, and ii)~it provides a heterogeneous pool of seed websites, which we subsequently use to expand our collection of search keywords. %
To ensure this, we curate scam websites from three different manually curated sources, contributing to the diversity of our dataset (details provided in Appendix~\ref{appendix:data_scam}).
The initial dataset of scam websites was compiled during March 2024.
Following the collection, we performed a preprocessing step to filter out domains that were either non-resolving or currently parked~\cite{alrwais2014understanding,vissers2015parking}.
Specifically, we employed a regular expression-based filter to inspect the HTML content of landing pages and identify parked domains. 
This filtering step ensures that only live scam websites are retained for downstream processing, as our system pipeline depends on features that require interaction with active scam websites.
Upon filtering, we end up with a total of 1,663 websites as part of the scam dataset.

\subsection{Benign websites}
\label{sec:dataset_benign}

Curating a dataset of benign websites is essential to train an accurate and reliable oracle classifier.
Unfortunately, this type of dataset is not directly available from previous work~\cite{bitaab2023beyond,kotzias2023scamdog}; therefore, we have to build one ourselves.
This task presents challenges.
While benign websites are significantly more prevalent than scam websites, the methodology used to sample benign domains has critical implications for the oracle's generalization capabilities, further affecting our downstream evaluation.
For example, sourcing benign examples exclusively from popular domain ranking lists such as Tranco or the Majestic Million may inadvertently bias the classifier toward established, high-traffic websites, potentially leading the oracle to fail generalizing to less popular but still legitimate websites. 
Moreover, it is crucial to ensure that the benign dataset mirrors the categorical diversity present in the scam dataset, to help the oracle distinguish between legitimate and fraudulent content within the same business vertical.
This balanced representation enhances the classifier's robustness and reduces the risk of category-specific overfitting.

To construct a systematic and representative benign dataset, we adopt a sampling strategy grounded in real-world user categorization.
Specifically, for each scam website filtered as an active scam in Section~\ref{sec:dataset_scam}, we query its domain on Trustpilot\footnote{https://www.trustpilot.com/}, a crowdsourced consumer-review platform, to determine whether the website has an associated profile or review history. 
For domains with a presence on Trustpilot, we extract the corresponding business category assigned by the platform.
This community-driven categorization scheme enables us to identify the business vertical of scam websites and subsequently guide the sampling of category-matched benign examples.

Out of 1,663 scam websites sourced from Section~\ref{sec:dataset_scam}, 592 websites had active profiles or reviews on Trustpilot, resulting in a total of 292 categories.
Using this set of inferred categories, we proceed to sample the 40 most \textbf{``relevant''} websites per category.
According to Trustpilot’s ranking criteria, ``relevance'' is determined by a combination of TrustScore, review volume, and business engagement.
In particular, qualifying businesses must actively solicit reviews and receive at least 25 verified reviews within the preceding 12 months.
Each API call to Trustpilot’s category returns 20 businesses (websites), and we make 2 API calls for each category to make the data collection more manageable. 
Limiting the highly ranked results (within the top 40 websites per category) also reduces the likelihood of any scam websites contaminating the benign set compared to including a larger set of results.
We regard this as a high-precision proxy for identifying reputable and trustworthy websites, thereby mitigating the risk of introducing mislabeled benign examples.
Following this procedure, we curate a total of 9,800 unique benign websites for inclusion in the training corpus.
To reflect the real-world distributional imbalance between scam and benign websites, we intentionally allow the dataset to be skewed toward benign examples rather than enforcing a strict class balance.
In the end, we have 1,663 scam websites and 9,800 benign websites spanning 292 diverse Trustpilot categories.

\subsection{Google Ads Keyword Suggestions}
\label{sec:dataset_keywords}

The third type of dataset utilized in our study comprises search queries associated with the scam domains.
While search queries can be either single tokens (e.g., ``shoes'') or multi-word expressions (e.g., ``cheap nike shoes''), we adopt the definition of a ``keyword'' throughout the rest of this work to represent both types of search queries.
To collect these keywords, we leverage the Google Ads Keyword Planner API\footnote{https://business.google.com/us/ad-tools/keyword-planner/}, which provides keyword suggestions based on either seed keywords or input domains.
Our work is the first to leverage Google Ads Keyword API to extract keywords for query sampling, as prior research mostly focused on static NLP methods such as TF-IDF~\cite{srinivasan2018exposing} and topic modeling~\cite{liu2023understanding}.
This deviation from static methods followed in previous works allows us to leverage Google's knowledge of large-scale search behavior, allowing for the generation of keyword suggestions grounded in real user search intent, historical query volume, and contemporary trends.
This enables a richer and more realistic representation of the types of search queries that might lead users to scam websites.

We use the Keyword Planner API to query each of the scam domains curated in Section~\ref{sec:dataset_scam}, retrieving keyword suggestions for every active scam website.
In addition to the keyword suggestions themselves, the API provides rich metadata for each suggestion, including estimated monthly search volume, competition level, and suggested bid price for a keyword in the context of the advertisement marketplace.
We further incorporate some of these attributes in the design and evaluation of keyword selection heuristics in Section~\ref{sec:sampling_attribute}.
Following this procedure, we obtain a total of approximately 1.5 million unique keyword suggestions originating from the 1,663 scam domains.

\subsection{SERP Collection}
\label{sec:dataset_serp}
The final type of dataset collection includes Search Engine Results Pages (SERPs) returned as results of querying the keyword suggestions obtained via the Google Ads Keyword Planner API.
For each query, we collect the list of web pages returned by different search engines.
We collect SERP data from four major search engines: i)~Google Search, ii)~Bing Search, iii)~Baidu Search, and iv)~Naver Search.
This multi-platform approach ensures diversity in the indexed content and ranking algorithms, thereby strengthening the generalizability of our findings.
We utilize the DataForSEO API\footnote{https://dataforseo.com/apis/serp-api} to query and retrieve SERP metadata for each candidate keyword.
The API simulates search queries using a clean user profile, i.e., without personalization or historical search context, and can be configured with a fixed geographic location (e.g., New York, USA)\footnote{https://docs.dataforseo.com/v3/serp-overview/}.
This methodology ensures that the returned SERP results reflect an unbiased snapshot of each engine’s ranking behavior at a specified time, unaffected by factors such as user history, geolocation-based personalization, or recommendation algorithms.
It also ensures that our data collection is conducted ethically, avoiding aggressive crawling of live search engine websites.

\section{Oracle}
\label{sec:oracle_model}

To calculate the toxicity and expansion of search queries, we need an oracle that can accurately determine whether a web page is a scam.
This has been common practice in similar research, for example, to identify websites that carry out drive-by-download attacks~\cite{invernizzi2012evilseed}.
Unfortunately, previous systems that are designed to detect scam websites were not made publicly available in a directly usable manner~\cite{bitaab2023beyond,kotzias2023scamdog}, and therefore cannot be used off-the-shelf as an oracle.
We therefore decide to implement our own oracle, building upon a diverse set of features proposed by previous works~\cite{bitaab2023beyond,kotzias2023scamdog} to enhance the diversity and coverage of features and produce a more accurate oracle.
The developed oracle helps us compute the toxicity and expansion scores for each search query, which we formally define as:
\[
\text{Toxicity} \;=\; \frac{\text{\# Websites flagged as scam by oracle}}{\text{\# Websites returned by a search query}}
\]
Additionally, we formally define the expansion metric as:
\[
\text{Expansion} \;=\; {\text{\# Scam websites returned by a search query}}
\]

Our oracle takes a domain as input, extracts a set of 103 features related to the domain registration information, DNS configurations, and features derived from the content of the webpage. %

\descr{Oracle Features.}
We build a comprehensive feature set for detecting scam websites by integrating and extending the features proposed in prior works~\cite{bitaab2023beyond,kotzias2023scamdog}.
Our oracle leverages a total of 103 features, which are grouped into five broad categories, as detailed below.
This unified feature representation captures both structural and behavioral attributes of websites, and reflects the current best practices in data-driven scam detection.

\begin{itemize}

    \item \textbf{Domain Ranking Features:}
These features characterize a domain’s reputation, authority, and popularity across the web, primarily using signals derived from third-party ranking services and threat intelligence providers.
For example, \texttt{majestic\_refips} and \texttt{majestic\_refsubnets} quantify the number of distinct referring IP addresses and subnets, serving as indicators of backlink diversity.
Features such as \texttt{majestic\_tldrank} and other Majestic-derived scores capture citation and trust flow metrics, which are commonly used to assess the credibility of a domain based on its position within the web hyperlink graph.
The \texttt{tranco} feature corresponds to a domain’s ranking in the Tranco list~\cite{pochat2018tranco}, a robust alternative to Alexa and other ranking services, while the \texttt{cisco} feature reflects reputation or threat score from Cisco Umbrella.
Collectively, these features provide a quantitative estimate of a domain’s visibility and trustworthiness within the web ecosystem.

\item \textbf{DNS Features:}\\
DNS (Domain Name System) features focus on the records associated with a domain’s configuration.
These include boolean indicators such as \texttt{dns\_has\_mx} and \texttt{dns\_has\_cname}, which denote the existence of mail exchange and canonical name records, respectively.
Complementary count-based features such as \texttt{dns\_num\_mx}, \texttt{dns\_num\_cname}, and \texttt{dns\_num\_aaaa} quantify the number of corresponding entries.
Additional DNS record types considered include DNAME, HINFO, NS, RP, SOA, and TXT, each offering insight into various aspects of domain functionality and administrative control.
The feature \texttt{dns\_domain\_verification\_count} reflects the number of verification-related records (e.g., for ownership or service validation).
Collectively, these features provide a comprehensive view of a domain’s operational and security-relevant DNS configuration.
Usage of these features is in line with prior work analyzing security implications of TXT records, MX records, and presence of SPF and DMARC-based protection mechanisms~\cite{kotzias2023scamdog}.

\item \textbf{URL Features:}\\
URL-based features examine the structure and composition of the domain or URL itself.
Features such as \texttt{tld} and \texttt{cheap\_tld} capture the type and cost tier of the top-level domain, distinguishing between commonly used, reputable TLDs and those often associated with abuse due to their low registration costs.
The feature \texttt{domain\_subwords} measures the number of semantically meaningful components in the domain name, potentially identifying keyword stuffing and obfuscation.
We use the open source library WordNinja~\footnote{https://github.com/keredson/wordninja}, which uses a probabilistic model to split words using NLP based on English Wikipedia unigram frequencies, to segment a domain into subwords.
To the best of our knowledge, \texttt{domain\_subwords} represents a novel feature not previously employed in prior scam detection systems.
Lexical indicators such as \texttt{url\_has\_hyphen} and \texttt{url\_has\_digit} detect the presence of special characters or digits, which are frequently used in deceptive or algorithmically generated domains.
Finally, \texttt{url\_subdomain\_count} quantifies the depth of the subdomain hierarchy, with excessive subdomain use sometimes correlated with malicious behavior.
Together, these features provide lightweight yet informative signals for assessing domain legitimacy, and have been utilized in prior work~\cite{bitaab2023beyond}.

\item \textbf{Domain Registration Features:}\\
These features are derived from WHOIS records and capture attributes related to the domain’s registration history and registrar identity. Temporal features such as \texttt{domain\_age} and \texttt{time\_to\_expiry} indicate the longevity and expected lifetime of a domain, both of which have been shown to correlate with domain credibility~\cite{bitaab2023beyond,kotzias2023scamdog}
Categorical features such as \texttt{registrar\_name}, \texttt{is\_cheap\_registrar}, \texttt{registrar\_country}, and \texttt{registrant\_country} characterize the registrar and the geographic provenance of the registration.
The boolean feature \texttt{privacy\_protected} reflects whether the registrant’s information is obscured via privacy protection services, a tactic frequently used in malicious or deceptive domain registrations~\cite{bitaab2023beyond}
Additionally, \texttt{free\_email\_provider} identifies whether the domain was registered using an email address from a free email provider, which may be indicative of low-cost or anonymous registrations. 
Collectively, these features provide a detailed profile of domain ownership and registration practices, which are essential for assessing trustworthiness and potential risk associated with scam websites created as part of a scale operation

\item \textbf{Content Features:}\\
Content-based features capture the structural and semantic elements present on a website’s landing page, offering insight into how the site is constructed and how it engages with users.
These features include indicators of social media integration (e.g., \texttt{facebook\_profile\_linked}, \texttt{twitter\_profile\_linked}, etc.) and communication channels (e.g., \texttt{num\_mailto\_links}, \texttt{num\_phone\_links}). Additional structural attributes such as \texttt{num\_links}, \texttt{num\_h1\_h6\_tags}, and \texttt{num\_css\_classes} quantify the complexity and layout of the HTML content.
Trust and compliance signals are also captured via features such as \texttt{review\_system\_linked}, \texttt{has\_app\_store}, \texttt{has\_review\_widget}, and \texttt{trustpilot\_present}, which identify the presence of third-party review platforms (e.g., Yelp, Trustpilot, etc.) or mobile application references.
Moreover, features like \texttt{presence\_work\_with\_us\_link} and \texttt{presence\_cookie\_consent\_notice} reflect the inclusion of user-facing transparency and recruitment information.
Collectively, these features provide a multifaceted view of the website's content, helping to infer its legitimacy, user engagement strategy, and adherence to common web practices associated with trustworthy entities.
These features capture particularly important signals for the oracle, as prior work on detecting fraudulent e-commerce platforms~\cite{kotzias2023scamdog} demonstrated that content-based features accounted for 5 of the top 10, and 10 of the top 15 most influential features in their machine learning–based classifier.
This highlights the strong predictive utility of content-centric attributes in distinguishing scam websites from legitimate ones.
\end{itemize}

Note that the content features captured as part of the oracle are based on the content present in the final destination of a URL loading chain.
Past studies~\cite{leontiadis2011measuring,leontiadis2014nearly} observed that malicious actors can compromise high-ranking benign websites and dynamically redirect traffic to illicit websites (i.e., Search Result Redirection attack), and such behavior can distort classification from the oracle and subsequent toxicity measurements.
The crawler we use for capturing these features overcomes any intermediate redirections and captures the features from the final destination, after the content of the page is loaded (including all the JavaScripts).
Subsequently, the oracle classifies a page exclusively on the final landing page and its associated eTLD+1.
This ensures that classification of discovered websites as scam / benign, and furthermore, the toxicity / expansion-related metrics are unaffected by the Search Result Redirection phenomenon.
A summary of the features is presented in Table~\ref{tab:feature_categories} (in Appendix).
These features are then passed to a supervised learning model trained to distinguish scam websites from benign ones.

\descr{Oracle Validation.}
To identify the most effective model for our oracle, we train and evaluate five different machine learning classifiers.
The models we evaluate include: i)~Random Forest, ii)~Gradient Boosting, iii)~Logistic Regression, iv)~Linear Discriminant Analysis Classifier, and v)~Support Vector classifier.
We evaluate all models using 5-fold cross-validation, employing an 80-20 stratified train-test split within each fold, preserving the class distribution reflecting real-world proportion of scam websites and benign websites.
A summary of the results from the 5 models is presented in Table~\ref{tab:oracle_eval}.
We find that the Gradient Boosting method performs the best in the task, with the highest F1 score, while the Random Forest model ranking a close second.
Across all folds, both precision and recall exceed 90\%, underscoring the robustness and reliability of the oracle in distinguishing scam websites from legitimate ones.
These findings are consistent with prior work, e.g., Scamdog Millionaire~\cite{kotzias2023scamdog} employed Random Forest as the classifier of choice for scam detection, further validating the effectiveness of ensemble-based models for this task.
In the rest of our experiments, we adopt the Gradient Boosting classifier as the implementation of our oracle. 

\begin{table}[H]
\centering
\begin{tabular}{|l|l|l|l|}
\hline
\textbf{Model}               & \textbf{Precision} & \textbf{Recall} & \textbf{F1 Score} \\ \hline
Random Forest                & 0.940              & 0.940           & 0.939             \\ \hline
\textbf{Gradient Boosting}            & \textbf{0.983}              & \textbf{0.965}           & \textbf{0.978}             \\ \hline
Logistic Regression          & 0.868              & 0.867           & 0.868             \\ \hline
Linear Discriminant Analysis & 0.933              & 0.932           & 0.932             \\ \hline
Support Vector Classifier    & 0.825              & 0.816           & 0.817             \\ \hline
\end{tabular}
\caption{ML model performance for scam detection.}
\label{tab:oracle_eval}
\end{table}

\section{Motivation of \systemname}
\label{sec:motivation}
The goal of \systemname is to retrieve a high proportion of scam websites surfaced through search results.
To the best of our knowledge, no prior work has conducted an empirical evaluation on identifying the most effective strategy for sampling keywords to support guided, search-based discovery of scam websites.
In this section, we evaluate the ability of query intent-based and search traffic-based attributes in surfacing search results with high toxicity, showing that they do not generalize across domains.
This is critical to develop a robust system, as scam types (and the language used by miscreants) change over time, also as a reaction to detection systems.
In this section, we first discuss the filtering steps that we apply to our query selection process in Section~\ref{sec:branding}.
We then discuss how we group scam categories for our evaluation in Section~\ref{sec:categories}.
Next, we present our sampling techniques, including traffic level, query intent-based metadata (Section~\ref{sec:sampling_attribute}), and semantic attributes derived from query understanding (Section~\ref{sec:sampling_segment}).
Finally, we present our evaluation of the two strategies in Section~\ref{sec:validation_dataset}, showing that they are insufficient, motivating the need for \systemname.

\subsection{Branded Keywords Filtering}
\label{sec:branding}

Query understanding literature distinguishes search queries between \emph{branded} (those including the name of a company, brand, or product line, e.g., ``nike air max'') and \emph{un-branded} (those containing more generic descriptions, e.g., ``trail running shoes'')~\cite{broder2002taxonomy}.
This distinction is particularly important in the context of scam website discovery, as branded queries tend to return SERPs dominated by legitimate and high authority domains associated with the brand, thereby lowering the toxicity of the keywords.
Consequently, filtering branded keywords is a critical step for both ensuring high toxicity keyword selection and maintaining focus on queries that are more likely to lead to scam websites.
For this reason, we filter the results suggested by the Google Keyword API to only keep un-branded queries.
We leverage the zero-shot text classification capabilities of Large Language Models (LLMs) (i.e. FLAN-T5-XXL) to build a branded keyword classifier.
Since this filter module is useful to clean up our dataset but is not a novel research contribution, we report its details in the Appendix~\ref{appendix:branded_classifier}.
The FLAN-T5-XXL model used as part of this module achieves a F1 score of 90\% on the test dataset, which we consider sufficiently accurate for the purposes of filtering branded keywords within our pipeline.
Following the validation of our branded keyword classification model, we apply it to the large-scale keyword dataset obtained in Section~\ref{sec:dataset_keywords}.
This filtering step results in a final corpus of approximately 1.2 million un-branded keywords, which we use for our experiments in the rest of the paper.

\subsection{Categories Selection}
\label{sec:categories}

The scam websites collected in Section~\ref{sec:dataset_scam}, and the category-matched benign websites, map to 292 unique Trustpilot categories.
To manage the size of keyword analysis and downstream evaluation, we focus on categories that are associated with at least five distinct scam domains. 
This leads us to a total of 49 different categories, which serve as the basis for all subsequent analyses in the rest of the paper.
To facilitate a more structured evaluation, we further group these 49 categories into 10 broader scam types, as outlined in Table~\ref{tab:categories}.
These aggregated scam types enable us to analyze the performance of keyword sampling strategies across semantically coherent business and service verticals.

\begin{table}[t]
  \scalebox{0.95}{
    \centering
\begin{tabular}{|l|l|l|}
\hline
\textbf{Scam Type} & \textbf{\# Cats} & \textbf{Example Categories (Scam Count)}         \\ \hline
Shopping / Fashion (\textbf{C1}) & 8                      & clothing, jewelry, beauty (363)         \\ \hline
Crypto / Money  (\textbf{C2})    & 9                      & cryptocurrency, finance, investment (278) \\ \hline
Adults / Gambling (\textbf{C3}) & 4                      & dating services, gambling (15)           \\ \hline
Medical / Pharmacy (\textbf{C4}) & 2                     & pharmacy, health (30)                    \\ \hline
Pets / Animals   (\textbf{C5})  & 2                     & pet stores, animals (44)                 \\ \hline
Electronics (\textbf{C6})       & 3                      & internet, phone (271)                     \\ \hline
Business / Admin (\textbf{C7})  & 3                      & business services, admin services (69)   \\ \hline
Education  (\textbf{C8})        & 3                      & career services, education training (39) \\ \hline
Marketing / Sales (\textbf{C9}) & 2                      & internet marketing, sales (36)           \\ \hline
Online Marketplace (\textbf{C10}) & 3                     & auction services, marketplace (114)      \\ \hline
\end{tabular}
  }
\caption{Distribution of different categories.}
\label{tab:categories}
\end{table}

\subsection{Query Sampling by Attributes}
\label{sec:sampling_attribute}
To better characterize the toxicity and expansion levels associated with different types of keywords, we study keyword sampling from the lens of: i)~search traffic-based features derived from Google's advertising ecosystem metadata and ii)~semantic attributes inferred using Natural Language Understanding (NLU) of the search queries~\cite{jansen2008determining}.
First, we leverage search traffic metadata provided by the Google Ads Keyword Planner API, which offers a real-world estimation of keyword popularity proxied via advertiser competition. 
Each keyword suggestion returned by the Google Ads Keywords API is labeled as: i)~Low competition, ii)~Medium competition, and iii)~High competition.
For example, a generic and high-value query such as ``nike shoes'' is typically labeled as High competition, due to strong brand recognition and commercial value.
In contrast, more specific or niche queries such as ``cheap sports shoes with durability'' tend to fall under the Medium or Low competition categories.
Prior research studying TSS websites~\cite{srinivasan2018exposing} found that approximately 91\% of the search phrases associated with TSS scams are labeled as either Low or Medium competition by the Google Ads Keyword Planner API. 
Furthermore, the study observed an inverse relationship between keyword popularity and scam toxicity (formally defined as \textit{pollution level} in their study), indicating that higher popularity terms tend to correspond with lower toxicity levels.
Based on these findings, we hypothesize that Low and Medium competition keywords are more effective for discovering highly toxic keywords and restrict our analysis to keywords labeled as Low or Medium competition throughout this study. 

Next, we classify keywords based on their semantic intent, using a taxonomy derived from established practices in query understanding~\cite{jansen2008determining}.
Specifically, we group keywords into the following intent-based categories using a publicly available multi-label query classification model\footnote{https://huggingface.co/dejanseo/Intent-XL}: 
\begin{itemize}
    
    \item \textit{Informational Intent Keywords:} Queries reflecting shoppers intent to gather more information or conduct preliminary research before buying the products and services.
    Recognizing this browsing intent of shoppers, e-commerce platforms commonly supplement their product catalogs with auxiliary content—such as blogs, buying guides, and customer testimonials—to enhance credibility, improve search visibility, and drive upstream traffic through informational queries (e.g., ``chosing right mattresss size,'' ``difference between etf and stocks'').

    \item \textit{Commercial Intent Keywords:} Queries reflecting a strong user intent to make a purchase or engage in transactional behavior.
    These types of queries often involve product comparisons, evaluations of alternatives, or searches for deals and discounts, thereby signaling a readiness to buy (e.g.,  ``best trail running shoes,'' ``dyson airwrap vs revlon styler'').
    
    \item \textit{Long Tail Keywords:} Queries that are highly specific multi-word search phrases, which often signal strong transactional or purchase intent.
    We define long tail keywords as search queries containing more than 3 words separated by a tokenizer. 
    These types of queries generally exhibit lower overall search volume, but they are associated with higher conversion likelihood due to their specificity and alignment with user intent (e.g., ``affordable standing desk with drawers,'' ``organic cotton baby clothes for sensitive skin.'')
\end{itemize}

\subsection{Query sampling by Segments}
\label{sec:sampling_segment}
Next, we explore segment-wise query sampling, where query segments potentially reflect behavioral or linguistic cues associated with deceptive scam-attracting tendencies.%
This allows for a systematic approach to sampling toxic queries, ultimately improving the discovery pipeline for scam websites.
We decompose search queries using Natural Language Understanding (NLU) techniques to analyze the relationship between different query segments (e.g., named entities, action verbs, price-related tokens,  etc.) and their associated toxicity scores.
Prior work in scam detection has focused on curating these query segments manually, and specific to a scam category~\cite{muzammil2025poorest}, suffering from limitations in coverage and temporal adaptability with shifts in the behavior of scam operations.
Automatically extracting query segments across multiple scam types presents an important avenue for detection as the diversity and volume of scam content evolve rapidly.
The motivation of this approach is to automatically identify the use of scare tactics, urgency, or persuasive calls to action as linguistic markers commonly employed by scam websites~\cite{bitaab2020scam,edwards2017scamming}.

Given a search query such as ``cheap christmas shoes for sale,'' a query segmentation model extracts semantically meaningful segments such as price (e.g., cheap), occasion (e.g., christmas), product (e.g., shoes), and modifiers (e.g., for sale).
This is a non-trivial problem that intersects with the broader task of query understanding~\cite{hagen2012towards}.
In this work, we adopt the approach recently introduced in QueryNER~\cite{palen2024queryner}, which formulates the query segmentation task as a Named Entity Recognition (NER) task.
The authors construct a domain-specific ontology over a large-scale shopping queries dataset~\cite{reddy2022shopping}, defining 17 distinct semantic categories for query segments.
These categories encompass a wide range of attributes, including product attributes (e.g., quantity, shape, color, condition), product metadata (e.g., department, occasion, origin), and shopping behavior indicators (e.g., price cues, transactional modifiers).
Among the various token categories identified, \textbf{modifiers} are of particular relevance to keyword toxicity, as they frequently serve to constrain, specify, or disambiguate user intent. 
For example, two different modifier terms ``without verification'' and ``trusted seller'' appended to a base query ``buy bitcoins,'' serve as different types of soft constraints that can influence both the type of e-commerce websites returned in the results and the likelihood of the query surfacing scam websites.
We use the publicly available BERT model \footnote{https://huggingface.co/bltlab/queryner-bert-base-uncased} fine-tuned on a training dataset of 7,000 queries and category tokens to extract semantic segments from input queries.
The model outputs token-level annotations, which we aggregate into coherent segments by post-processing the output.
These segments are subsequently used as heuristics for sampling queries based on the presence of each token category.

\subsection{Validation}
\label{sec:validation_dataset}
To systematically evaluate the effectiveness of different keyword segment and attribute-based methods for sampling the most \textbf{``toxic''} search queries, we construct a targeted validation dataset.
Given the computational and monetary cost (through API costs) associated with querying SERP results for over 100,000 un-branded candidate keywords, we sample a representative subset of keywords for validation.
Specifically, we construct a targeted validation dataset by randomly sampling a fixed number of keywords (100 per scam category), stratified across the keyword attributes and query segments discussed in the aforementioned sections.
We restrict SERP data collection for this dataset to three major search engines: i)~Google, ii)~Bing, and iii)~Baidu to balance coverage and efficiency.
In contrast, our large-scale evaluation in Section~\ref{sec:section_wild} will use SERP data from all four search engines for broader generalizability.
Sampling 100 keywords per category across 49 scam-relevant categories yields a total of 4,900 unique search queries, from which we collect approximately 125,000 unique websites via SERP responses to serve as the validation dataset for assessing the effectiveness of various heuristic strategies in surfacing scam websites.
We compute the toxicity and expansion scores for each of the 4,900 keywords as the proportion of websites returned by the keyword that are classified as scams by the oracle.
For each keyword, we use all the SERP results returned by the DataforSEO API, deduplicated by root domain, to compute the toxicity and expansion scores (85 results on average per query).
We do not limit the toxicity calculation to any top results in this step to get an overall estimate of how ``toxic'' the set of results returned by a query is, regardless of how ``top'' ranked the scam websites themselves are.
While it is useful to consider where the websites are being ranked in terms of users being exposed to the websites in real-world SERP results, we don’t filter the SERP results by position for the toxicity estimation and modeling task, since for the purpose of a proactive detection system, our objective is to identify as many scam websites surfaced by a query as possible.
The prevalence of scam websites within a category is highly category dependent (e.g., the cryptocurrency category exhibits a significantly higher susceptibility to scams than categories such as ``education and training.'')
To account for this, we analyze the effectiveness of the sampling strategies within individual categories, rather than aggregating results across all categories.
Specifically, we select C1, C2, C3, C4, and C5 as the evaluation categories, as they correspond to scam types most prominently victimizing real-world users, according to recent findings by Kotzias et al.~\cite{Kotzias2025CtrlAltDeceiveQU}.
This selection grounds our evaluation in realistic, high-impact scam categories, thereby enhancing the practical relevance and external validity of our results.

\descr{Toxicity by query attribute.}
We evaluate the effectiveness of sampling keywords using different attributes by analyzing their relationship with query-specific toxicity scores.
For each attribute heuristic introduced in Section~\ref{sec:sampling_attribute}, we filter the subset of keywords that match the attribute (e.g., query is of informational intent, query is of long tail distribution etc.)
To obtain robust estimates of the average toxicity and expansion scores associated with each attribute, we employ a bootstrapped sampling procedure over the respective scores of the filtered query subsets.
Specifically, for each query attribute, we conduct 1,000 bootstrap simulations by randomly sampling 20 toxicity scores (with replacement) from the corresponding subset of matched queries and compute their mean.
This approach yields an empirical distribution of average toxicity values, enabling us to get an estimate of how strongly each attribute correlates with query toxicity.
We repeat a similar process for estimating the query expansion scores.

\begin{table}[t]
  \centering
  \resizebox{0.5\textwidth}{!}{%
    \begin{tabular}{lcc|cc|cc|cc|cc}
      \toprule
      & \multicolumn{2}{c}{\textbf{C1}}
      & \multicolumn{2}{c}{\textbf{C2}}
      & \multicolumn{2}{c}{\textbf{C3}}
      & \multicolumn{2}{c}{\textbf{C4}}
      & \multicolumn{2}{c}{\textbf{C5}} \\
      \cmidrule(lr){2-3}\cmidrule(lr){4-5}\cmidrule(lr){6-7}\cmidrule(lr){8-9}\cmidrule(lr){10-11}
      \textbf{Sampling Type}
       & \textbf{Tox.} & \textbf{Exp.}
       & \textbf{Tox.} & \textbf{Exp.}
       & \textbf{Tox.} & \textbf{Exp.}
       & \textbf{Tox.} & \textbf{Exp.}
       & \textbf{Tox.} & \textbf{Exp.} \\ 
      \midrule
      \textbf{Max}
       & 0.20 & 66.05  
       & 0.23 & 79.75  
       & 0.28 & 94.15  
       & 0.39 & 133.65  
       & 0.274 & 92.95  \\ 
      \midrule[\heavyrulewidth]  %
      Informational      
       & 0.097 & 32.60  & 0.071 & 24.20  & 0.089 & 30.33  & 0.147 & 49.33  & 0.144 & 45.22  \\ 
      Commercial         
       & 0.071 & 23.83  & 0.099 & 33.48  & 0.130 & 43.72  & 0.092 & 31.11  & 0.117 & 39.86  \\ 
      Low Competition    
       & 0.116 & 38.72  & 0.097 & 32.96  & 0.133 & 45.04  & 0.209 & 59.96  & 0.191 & 64.64  \\ 
      Medium Competition 
       & 0.121 & 40.84  & 0.095 & 32.12  & 0.129 & 43.24  & 0.165 & 55.63  & 0.175 & 59.39  \\ 
      Long Tail          
       & 0.081 & 27.47     & 0.094 & 32.06     & 0.129 & 43.86  & 0.194 & 64.64  & 0.093 & 31.62  \\ 
      \bottomrule
    \end{tabular}%
  }
  \caption{Performance of attributes for keyword toxicity and expansions.}
  \label{tab:toxicity_attribute}
\end{table}

Table~\ref{tab:toxicity_attribute} reports both the mean toxicity scores and expansion rate associated with queries filtered by different attributes across the evaluation scam categories (C1–C5).
We also provide a reference \emph{Max} score for each category, providing the upper bound of average keyword toxicity and expansion when the keywords are sampled by sorting their ground truth toxicity and expansion scores.
The results illustrate that certain attributes are more consistently associated with elevated toxicity and expansion levels.
In particular, queries characterized by Low Competition and Medium Competition demonstrate higher average toxicity and expansion across all categories, with Low Competition achieving the highest toxicity in C4 (0.209) and Medium Competition following closely in C5 (0.175).
This finding is consistent with prior work on TSS; however, while previous analysis was specific to only one category of scam, we find this relationship generalizes across a broader set of scam categories.
Moreover, upon referencing the performance of these attributes to the \emph{Max} score, we find that even the best attribute-based sampling provides largely suboptimal results in terms of the toxicity and expansion, across all 5 categories.

\descr{Toxicity by query segments.}
\label{sec:results_segments}
Next, we assess keyword toxicity and expansion associated with sampling strategies based on query segments.
We treat segment types introduced in Section~\ref{sec:sampling_segment} (e.g., price, modifier, etc.) as semantic attributes that capture linguistic patterns or tactics commonly employed by scam websites.
While the taxonomy proposed by ~\cite{palen2024queryner} defines 17 different token types, not all of them are relevant or provide sufficient coverage across scam categories.
We observe that only five token types: i)~Core Product Type, ii)~Content, iii)~Product Name, iv)~Modifier, and v)~Price consistently meet the minimum threshold of at least 20 keyword matches per category, allowing for stable bootstrapped toxicity estimation. 
Therefore, we focus our analysis on these five high-coverage query segments, as they offer both semantic interpretability and robust empirical support for evaluating segment-based query sampling strategies across diverse scam categories.
For each scam category (C1 through C5), we identify the top 20 highest-ranked query segments within each token type (e.g., modifier, price, etc.) and filter the set of keywords that contain any of these segments.
For multi-word segments (e.g., ``for sale,'' ``freshly used'', etc.), a keyword is considered a match if all constituent words of the segment appear anywhere within the query text.

Table~\ref{tab:toxicity_segments} presents the mean toxicity and expansion scores for each token type across the five scam categories (C1–C5).
Similar to the previous section, the results reveal that all query segments consistently produce suboptimal sampling in reference to the \emph{Max} scores, and no single query segment type emerges as uniformly effective across all scam categories. %
The variability observed in Table~\ref{tab:toxicity_segments} further illustrates the generalization challenges associated with segment-based, attribute-based, or any rule-based keyword sampling methods.
In the absence of category-specific priors, query segment-based keyword sampling approaches become suboptimal.%
This limitation is particularly critical in the proactive discovery of emerging scam categories, where contextual signals necessary for selecting relevant segments or attributes are often absent.
These findings reinforce the limitations of static heuristics and highlight the need for a more robust and adaptive keyword discovery strategy, as embodied by the data-driven approach proposed in \systemname.

\begin{table}[t]
  \centering
  \resizebox{0.5\textwidth}{!}{%
    \begin{tabular}{l cc cc cc cc cc}
      \toprule
      & \multicolumn{2}{c}{\textbf{C1}}
      & \multicolumn{2}{c}{\textbf{C2}}
      & \multicolumn{2}{c}{\textbf{C3}}
      & \multicolumn{2}{c}{\textbf{C4}}
      & \multicolumn{2}{c}{\textbf{C5}} \\
      \cmidrule(lr){2-3}
      \cmidrule(lr){4-5}
      \cmidrule(lr){6-7}
      \cmidrule(lr){8-9}
      \cmidrule(lr){10-11}
      \textbf{Sampling Type}
       & \textbf{Tox.} & \textbf{Exp.}
       & \textbf{Tox.} & \textbf{Exp.}
       & \textbf{Tox.} & \textbf{Exp.}
       & \textbf{Tox.} & \textbf{Exp.}
       & \textbf{Tox.} & \textbf{Exp.} \\
      \midrule
      \textbf{Max}
       & 0.20 & 66.05  
       & 0.23 & 79.75  
       & 0.28 & 94.15  
       & 0.39 & 133.65  
       & 0.274 & 92.95  \\
      \midrule[\heavyrulewidth]
      Core Product Type
       & 0.09 & 33.83
       & 0.09 & 30.87 
       & 0.14 & 48.03
       & 0.22 & 74.99
       & \textbf{0.13} & 43.97 \\
      Content
       & 0.08 & 47.08
       & 0.09 & 32.54
       & 0.18 & 61.03
       & 0.23 & 79.69
       & 0.08 & 28.87 \\
      Product Name
       & 0.08 & 27.23
       & 0.09 & 31.84
       & 0.14 & 48.75
       & 0.13 & 45.31
       & 0.12 & 41.27 \\
      Modifier
       & 0.09 & 32.72
       & 0.10 & 34.58
       & 0.16 & 54.92
       & 0.20 & 69.04
       & 0.08 & 28.98 \\
      Price
       & 0.08 & 40.52
       & 0.08 & 27.72
       & 0.14 & 48.75
       & 0.24 & 80.86
       & —    & —    \\
      \bottomrule
    \end{tabular}%
  }
  \caption{Performance of entities for keyword toxicity.}
  \label{tab:toxicity_segments}
\end{table}

\begin{table}[ht]
  \centering
  \resizebox{0.5\textwidth}{!}{%
    \begin{tabular}{l c c c c c}
      \toprule
      \textbf{Source}
      & \textbf{C1}
      & \textbf{C2}
      & \textbf{C3}
      & \textbf{C4}
      & \textbf{C5} \\
      \midrule
      C1 & \textbf{0.09} & 0.08 (\textcolor{red}{–1\%})  & 0.13 (\textcolor{red}{–4\%})  & 0.18 (\textcolor{red}{–5\%})  & 0.10 (\textcolor{red}{–3\%}) \\ 
      C2 & 0.09          & \textbf{0.10}                 & 0.17                          & 0.21 (\textcolor{red}{–2\%})  & 0.09 (\textcolor{red}{–4\%}) \\ 
      C3 & 0.09          & 0.09 (\textcolor{red}{–1\%})  & \textbf{0.17}                 & 0.14 (\textcolor{red}{–9\%})  & 0.09 (\textcolor{red}{–4\%}) \\ 
      C4 & 0.08 (\textcolor{red}{–1\%}) & 0.10                & 0.10 (\textcolor{red}{–7\%})  & \textbf{0.23}                 & 0.10 (\textcolor{red}{–3\%}) \\ 
      C5 & 0.08 (\textcolor{red}{–1\%}) & 0.10                & 0.15 (\textcolor{red}{–2\%})  & 0.18 (\textcolor{red}{–5\%})  & \textbf{0.13}                \\ 
      \bottomrule
    \end{tabular}%
  }
  \caption{Cross Category Comparison Table.}
  \label{tab:category_comparison}
\end{table}

\descr{Cross-Category Generalization of Heuristic Selection.}
Our findings in the previous sections motivate an important question: Can heuristic signals derived from toxic query segments in one scam category generalize to others?
Enabling such generalization would greatly improve the effectiveness of query-powered systems to discover scam sites, removing the need for category-specific keywords and accounting for time and evasion-related changes.
In this question, we investigate whether high toxicity query segments identified in one category can generalize effectively for sampling toxic queries across other scam categories.
For example, in category C1, the query segments most associated with toxic keywords include terms ``cheap,'' ``free,'' and ``sale.''
We evaluate whether these segments retain their discriminatory power when applied to other scam categories.
For each scam category in the validation set, we extract the top 20 query segments most associated with high toxicity queries and compute the average toxicity of queries containing these segments, using the matching and scoring methodology previously outlined in Section~\ref{sec:results_segments}.
This provides an upper bound on the predictive capacity of in-category query segments.
We then apply these query segments to the remaining categories and measure their effectiveness in sampling toxic queries for other scam categories.
Specifically, using a reference category (e.g., C1), we use its top toxic query segments to sample queries in other categories (C2–C5) and compute their bootstrapped toxicity scores.%
This process is repeated for each category (C1 through C5) as the reference, resulting in a matrix of cross-category toxicity scores.

Table~\ref{tab:category_comparison} presents the results of our cross-category generalization experiment, where each row corresponds to the source category from which the reference query segments were extracted (denoted as ``Source''), and each column corresponds to the target category where the reference query segments were applied for sampling the queries.
The diagonal entries indicate within-category performance, while off-diagonal values reflect the toxicity scores obtained by applying query segment heuristics from the source category to sample queries in a different target category.
The red percentages indicate the relative drop in toxicity compared to the diagonal (i.e., in-domain) baseline.
Across all source categories (C1–C5), we observe a consistent drop in toxicity when the source heuristics are applied to other target categories.
For example, using C1’s top segments in C3, C4, and C5 results in a 4–5\% reduction in toxicity relative to C1's own performance.
This trend applies across all categories, as no source category provides a heuristic set that consistently generalizes to all others without incurringa  measurable performance drop.
These results indicate that, even when derived from labeled data, query segment–based heuristics exhibit limited generalization across scam categories.

\descr{Takeaways.}
Sampling keywords by heuristics such as query attributes and query segments performs suboptimally in identifying the most effective queries in terms of toxicity and expansion, and does not generalize across different categories.
This shows the need for a generalizable, data-driven method to achieve the objective of sampling the most toxic set of keywords for improved discovery of scam websites, which we present in the next section.

\section{\systemname}
\label{sec:system}

The empirical observations made in the previous section motivate the need for a scalable, data-driven framework that can learn to identify toxic queries, overcoming the limitations of category-specific heuristics and rules.
Guided by this insight, we now introduce our system \systemname, centered around modeling the toxicity of search keywords.
The objective of the model trained in \systemname is to estimate each query's underlying toxicity score without having access to its corresponding SERP results at test time.
Given a pool of candidate queries collected from Google Ads Keywords API, \systemname predicts a continuous score of keyword toxicity (between 0-1) for the queries. %
The highest-ranked queries are then selected and queried for search engine results, enabling a guided approach for the discovery of scam websites in the wild.

In this section, we first illustrate the modeling considerations behind \systemname.
We then present our evaluation strategy and discuss our system's implementation.
Finally, we present the results of \systemname on the toxicity prediction task, compare it against relevant baselines, and discuss the takeaways of our results.

\subsection{\systemname: Modeling Considerations}

Using query segment classification models like \textit{BrandNER} (which we previously used for attribute-level toxicity analysis in Section~\ref{sec:sampling_attribute}) is insufficient for \systemname, as these models are unable to capture the underlying intent and latent attributes influencing SERP-level toxicity of a query, which is at the core of our approach.%
\systemname needs a more comprehensive modeling framework that can learn these latent properties directly from real-world SERP data, rather than relying on intermediate lexical features.
For this reason, we adopt transformer-based models like DistilBERT~\cite{sanh2019distilbert} to predict query toxicity directly.
This model is ideal for our task, because it avoids the brittleness that is inherent to models based on intermediate lexical features, and can capture rich linguistic and structural features through its contextualized embeddings~\cite{jawahar-etal-2019-bert}.
Most importantly, these structural representations are learned in a task-specific manner during fine-tuning, aligning language understanding with the toxicity prediction objective.

A key innovation of \systemname is that rather than stopping at vanilla fine-tuning of transformer models over query-toxicity score pairs, we enhance the learning setup by integrating an additional modality of information: the SERP metadata associated with each query.
In the standard setup, the model learns to predict the latent toxicity of a query purely from its textual representation.
By incorporating SERP level metadata, we introduce auxiliary signals that help bridge the gap between the query and its observed toxicity score by offering insight into the nature of the search results that contributed to (or minimized) that toxicity.
These metadata capture the qualitative aspects of the retrieved content (e.g., webpage descriptions) that are not directly observable from the query alone, thereby grounding the abstract notion of toxicity in real-world search engine retrievals.
However, a key practical constraint in this setting is that while SERP data is available during training, it is unavailable at inference time, as querying search engines at scale is counterintuitive for real-time query sampling (and computationally expensive).
To overcome this issue, we adopt the Learning Under Privileged Information (LUPI) framework~\cite{vapnik2009new}, which posits that access to supplementary information during training, even if unavailable at inference time, can significantly improve the generalization ability of the model.
In our case, the SERP metadata available during training serves as privileged information that guides the model in learning more discriminative representations of query toxicity.%
This enables the model to establish a richer query-to-toxicity mapping, even when SERP information is not available at test time.

To operationalize the LUPI framework within a Machine Learning setting, we adopt a teacher–student distillation paradigm in which the teacher model has access to the SERP metadata (the privileged information) during training, and the student model learns to predict query toxicity using only the query text.
This approach is motivated by the intuition that effective teachers do more than simply providing the correct answer: they also convey the underlying rationale behind those answers~\cite{lambert2018deep}.
In our case, rather than asking the model to directly infer a continuous toxicity score from queries alone, we use the SERP metadata to provide additional context that explains how a particular query received its toxicity label.
The resulting student model is optimized to predict toxicity solely from the query text, with no access to SERP information at inference time.
Furthermore, the privileged SERP information serves as an inductive bias, enabling the regression model to more effectively approximate the underlying relationship between query semantics and toxicity.
Given access to SERP metadata for queries in the training dataset, we train our LUPI model using the teacher–student framework and evaluate the student model to predict toxicity scores using only query text, without issuing search queries.

\subsection{Leave-One-Category-Out Cross-Validation}
\label{sec:5foldcv}

\systemname's goal is to capture the latent notion of query toxicity in a category-agnostic manner.
Given the absence of prior datasets or standardized benchmarks for this task, we adopt a 5-fold cross-validation strategy to evaluate the generalization performance of our model.
To evaluate the model's generalization across scam types, we adopt a leave-one-category-out cross-validation setup (analogous to Leave-One-Out Cross-Validation (LOOCV) in standard ML), where each fold holds out queries from a distinct scam category as the test split.
In each fold, the model is trained on queries from four categories and evaluated on the held-out fifth category.
This setup ensures that the test queries originate from entirely unseen business verticals, providing a rigorous assessment of the model's ability to generalize beyond category-specific patterns.
This evaluation framework represents a \textit{hard generalization} setting, in which the model must learn to infer toxicity signals purely from the structure and semantics of the query, without relying on implicit correlations with previously seen queries from the same category.
By enforcing this separation, we mitigate risks of unintended feature leakage, reduce overfitting, and more faithfully assess the extent to which the model captures category-agnostic properties of query toxicity.

We construct five cross-validation folds, with each fold holding out one scam category for evaluation while training on the remaining four.
Consistent with the evaluation criteria outlined in Section~\ref{sec:validation_dataset}, we select C1, C2, C3, C4, and C5 as the held-out categories, allowing for fair and consistent comparison across baseline methods.
As an example of the Leave-One-Category-Out Cross-Validation setup, in Fold 1 (F1), we train on all categories except C1 (Shopping / Fashion) and evaluate on C1.
We repeat this process for the remaining folds.
Note that this leave-one-category-out cross-validation framework is flexible and can be readily extended to assess model performance on any other scam categories (C6-C10) or new scam categories identified by future studies, depending on specific research goals.

\subsection{\systemname: Implementation}

As discussed, the training procedure of \systemname follows a two-stage approach grounded in the Learning Under Privileged Information (LUPI) paradigm.
In the first stage, we train a teacher model that has access to both the query text and the corresponding privileged information in the form of SERP metadata.
This model is tasked with learning to predict query toxicity using the additional SERP modality.
Importantly, this step functions not only as the foundation for subsequent distillation but also as a validation of the utility of incorporating privileged information through the additional modality.
By comparing the teacher model’s performance to baseline transformer models (e.g., DistilBERT) that rely solely on query inputs, we can empirically assess the added value of incorporating SERP level metadata.
In the second stage, we then train a student model that has access only to the query text.
The student is optimized to mimic the teacher’s output, thereby distilling the additional knowledge encoded via the privileged SERP information into a model deployable under real-world constraints, where such metadata is not available at inference time.

\subsubsection{Teacher Model}
\begin{figure*}{}{}
\centering
\scalebox{0.8}{
    \includegraphics[width=\linewidth]{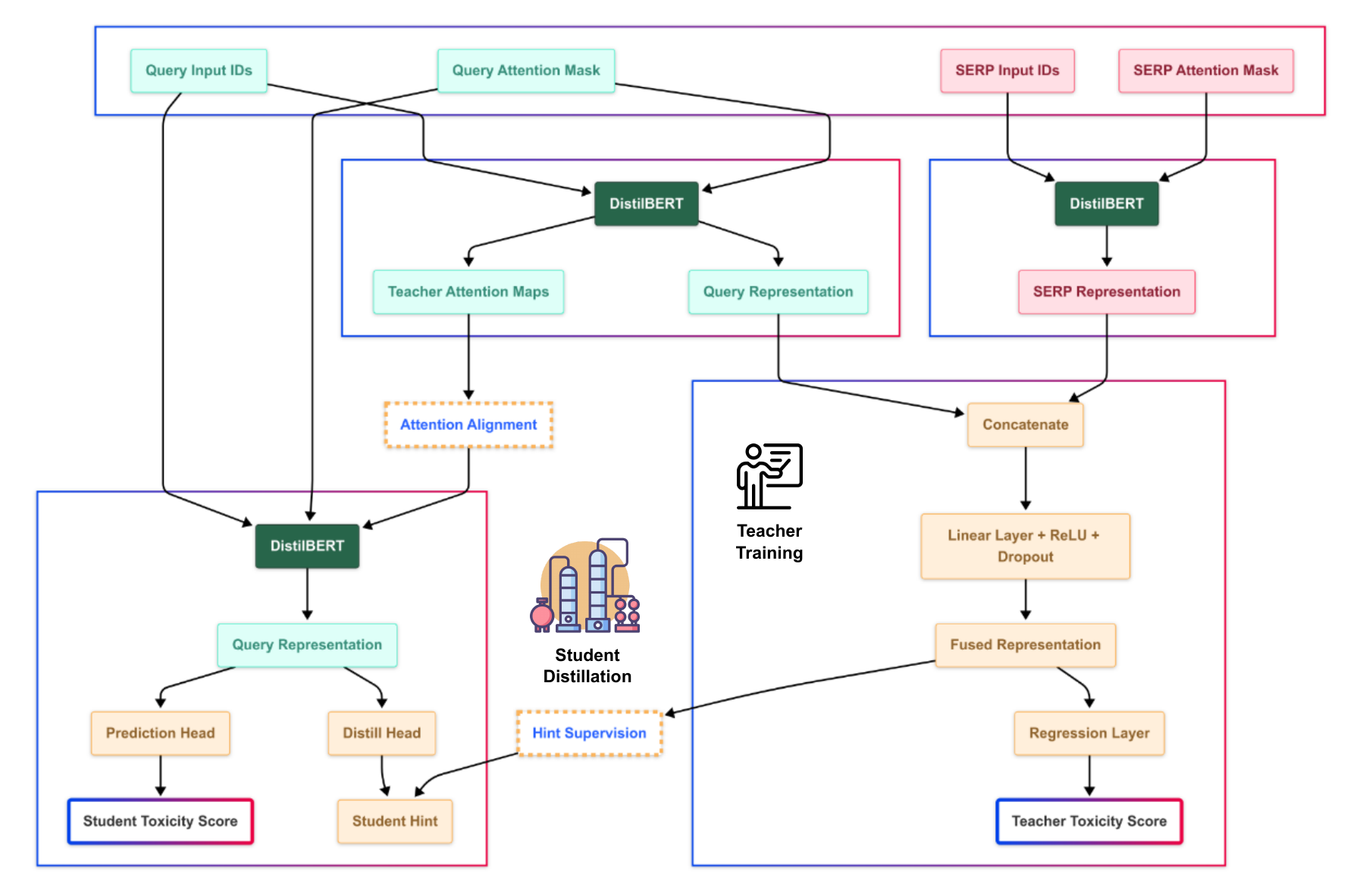}}
  \caption{Overview of \systemname pipeline.}
  \label{fig:lokipipeline}
\end{figure*}

The teacher model implemented in \systemname is designed to predict query toxicity based on the contextual relationship between a query and its corresponding Search Engine Results Pages (SERP).
To model these two distinct input modalities, the architecture incorporates two independent instances of the DistilBERT encoder: one dedicated to processing the query and the other to processing the SERP content.
This architecture enables the model to learn specialized contextual representations for each modality, without forcing a shared embedding space.
The query representation is extracted from the embedding of the special classification token (\texttt{[CLS]}) in the final hidden state of the query encoder.
This token is primarily used in transformer architectures such as BERT to summarize the entire input sequence for downstream classification or representation learning tasks~\cite{devlin2019bert}.
For SERP inputs that may consist of multiple result entries per query, we flatten the entries, encode each individually using the SERP encoder, and apply mean pooling across all entries to derive a single aggregated SERP embedding.
These two embeddings are then concatenated and passed through a fusion linear layer followed by ReLU activation and dropout for regularization.
The final output layer is a regression head that produces a continuous toxicity score.
Optionally, the model also returns the attention map from the query encoder, which is later utilized in the distillation process to guide the student model.
A visual representation of the teacher model architecture is presented in Figure~\ref{fig:lokipipeline}.

\subsubsection{Selecting the privileged information}
An important aspect of the teacher model implementation involves identifying the proper subset of the available SERP metadata to be used as the \textbf{\textit{privileged information}} during training.
Given the diversity and volume of metadata returned by search engines through the DataForSEO API, careful selection is required to determine the most informative signals for guiding the teacher model.
To this end, we explore multiple strategies for selecting privileged information across four distinct axes, each capturing a different facet of the SERP context, as detailed below.

\begin{itemize}
    \item \textbf{Search Engine:} We collect SERPs available for each query in the validation dataset from 3 different search engines (i.e., Google, Bing, and Baidu).
    We experiment with selecting the SERP from each of the search engines individually and together.
    
    \item \textbf{SERP Metadata:} The DataForSEO API returns a rich set of metadata for each SERP entry, including the page title, snippet description, presence in Q\&A forums, mentions in Google Business, and other contextual indicators.
    To determine which metadata fields are most effective as privileged information, we conduct ablation experiments using different subsets of these features.

    \item \textbf{SERP Filtering:} The domain of SERP (benign or scam websites) can affect the model training.
    To address this, we explore filtering strategies wherein only SERP entries corresponding to scam websites, as identified by our oracle labeling mechanism, are retained as privileged information.
    We compare this filtered setup to a baseline that includes all SERPs without filtering.

    \item \textbf{SERP Size and Ranking:} Finally, we investigate the impact of both the number of SERP entries encoded per query and the strategy used to select them.
    We compare two sampling strategies: random selection versus rank-based selection, where SERPs are chosen according to their position in the original search engine results.
    On average, each search query for a search engine returns around 85 SERPs.
    Additionally, we experiment with varying the number of SERP entries included per query (ranging from 5 to 50.)

\end{itemize}

We identified the optimal configuration for training the teacher model by systematically evaluating all four axes of privileged information selection in a grid search setup, varying the search engine source, metadata type, filtering strategy, and number and ranking of SERP entries.
Specifically, using the \textbf{top 20 ranked} SERP \textbf{descriptions} from \textbf{Google}, filtered to include \textbf{only scam domains}, yielded the best performance for the teacher model.
This configuration is adopted for all subsequent training and evaluation of the teacher models.

\descr{Teacher model results.}
The teacher model was trained using the AdamW optimizer~\cite{loshchilov2018decoupled} with a learning rate of 2e-5 for 5 epochs.
The objective function was the Mean Absolute Error (MAE), implemented via the L1 loss, to encourage robust regression of toxicity scores.
We select the best model based upon validation loss on a held-out validation dataset (sampled as 10\% of the test dataset for each fold).
Evaluation is conducted using the stratified 5-fold leave-one-category-out cross-validation setup described in Section~\ref{sec:5foldcv}.
As a baseline, we compare against the DistilBERT model trained solely on query text, with no access to privileged SERP metadata and the best-scoring query sampling strategies (both attribute-based and token-based from Section~\ref{sec:validation_dataset}).
Additionally, to approximate an upper bound on achievable toxicity scores within each category, we include the \emph{Max} score by sampling the top 20 queries based on their ground truth toxicity scores.

The evaluation results are summarized in Table~\ref{tab:eval_teacherserp}.
Across all five held-out categories, the teacher model incorporating SERP metadata consistently outperforms the baseline query-only model.
Notably, for categories C3 and C5, the teacher model’s performance approaches the upper bound in terms of both toxicity and expansion, providing an improvement over the suboptimal query sampling of attribute-based approaches found in Section~\ref{sec:validation_dataset}. 
These results provide strong empirical validation of our intuition: incorporating SERP metadata as privileged information significantly enhances the model’s ability to learn the latent toxicity of queries.

\begin{table}[t]
  \centering
  \resizebox{0.5\textwidth}{!}{
    \begin{tabular}{l *{10}{>{\normalsize}c}}
      \toprule
      & \multicolumn{2}{c}{\textbf{C1}}
      & \multicolumn{2}{c}{\textbf{C2}}
      & \multicolumn{2}{c}{\textbf{C3}}
      & \multicolumn{2}{c}{\textbf{C4}}
      & \multicolumn{2}{c}{\textbf{C5}} \\
      \cmidrule(lr){2-3}
      \cmidrule(lr){4-5}
      \cmidrule(lr){6-7}
      \cmidrule(lr){8-9}
      \cmidrule(lr){10-11}
      \textbf{Sampling Type}
        & \textbf{Tox.} & \textbf{Exp.}
        & \textbf{Tox.} & \textbf{Exp.}
        & \textbf{Tox.} & \textbf{Exp.}
        & \textbf{Tox.} & \textbf{Exp.}
        & \textbf{Tox.} & \textbf{Exp.} \\
      \midrule
      \textbf{Max}
        & 0.19 & 66.05 & 0.23 & 79.75 & 0.28 & 94.15 & 0.39 & 133.65 & 0.27 & 92.95 \\
      \midrule[\heavyrulewidth]
      \textbf{Best Attribute / Entity}
        & 0.12 & 40.84   & 0.10 & 34.58   & 0.18 & 61.03   & 0.24 & 80.86   & 0.19 & 64.64   \\
      \textbf{DistilBERT}
        & 0.09 & 32.16   & 0.17 & 54.52  & 0.23 & 75.36   & 0.26 & 98.33   & 0.20 & 68.32   \\
      \textbf{Teacher}
        & 0.15 & 53.45   & 0.19 & 67.3   & 0.26 & 84.85   & 0.30 & 112.55   & 0.25 & 84.00   \\
      \textbf{Student}
        & 0.12 & 50.34   & 0.20 & 67.9   & 0.24 & 79.45   & 0.30 & 110.75   & 0.22 & 75.95   \\
      \bottomrule
    \end{tabular}%
  }
  \caption{Performance of \systemname's teacher and student models with SERP information.}
  \label{tab:eval_teacherserp}
\end{table}

\subsection{Distillation of Student Model}
Next, we distill this representation learnt by the teacher model using a feature distillation framework.
The goal of this distillation process is to transfer the latent query-toxicity mapping from the privileged teacher to a student model that relies solely on query inputs, enabling the student model to predict toxicity scores of keywords in the wild, where SERP metadata is unavailable.

The student model is a lightweight regression architecture, structurally identical to the baseline model, designed to predict toxicity scores using only the query input.
Unlike the baseline, however, the student is trained to approximate the behavior of the teacher model while simultaneously learning to predict toxicity scores.
It is to note that, compared to model-based distillation, where a teacher model has higher complexity than the student model (in terms of parameter size, or neural complexity of the model itself), feature-based distillation can use student models that are of comparable capacity and complexity to the teacher models.

Consistent with the baseline and teacher SERP models, the student model uses a single instance of the DistilBERT transformer encoder to generate contextualized representations of the query.
Specifically, the final hidden state corresponding to the \texttt{[CLS]} token is extracted as the query level embedding.
This representation is then processed by two task-specific heads: i)~prediction head, and ii)~distillation head.
The first head, the prediction head consists of a dropout layer followed by a two-layer feedforward network with ReLU activation, resulting in a continuous output estimating the toxicity score.
In parallel, the distillation head consists of a projection layer that transforms the query representation into an intermediate representation intended to mimic the teacher model's fused representation, which was originally trained with access to the privileged information.
Note that this projection layer is a simple linear layer, as the teacher model and student model share the same architectural backbone, and this architectural symmetry eliminates the need for complex feature alignment typically needed in distillation between heterogeneous architecture~\cite{hao2023one}.
The output from the distillation head thus facilitates feature-based knowledge distillation, which we use as part of training the student model.
A visual representation of the student model architecture and the distillation process is presented in Figure 2.
To effectively transfer the teacher’s knowledge into the student model, we employ a multi-objective distillation framework comprising four complementary loss components:

\begin{itemize}
    \item \textbf{Ground Truth Loss:} Loss aligning with direct supervision using MAE between the student's prediction and the true label.
    \item \textbf{Prediction Matching Loss:} Loss aligning student predictions with the teacher's predicted scores using MAE.
    \item \textbf{Hint Matching Loss:} Loss to supervise the reconstruction of the student's intermediate representation as close as possible to the teacher's fused representation using Mean Squared Error (MSE).
    \item \textbf{Attention Matching Loss:} Minimizes the distance between the attention maps of corresponding layers in the student and teacher model using Mean Squared Error.
\end{itemize}

The loss function used during training the student model is formally represented as: 
\begin{equation}
\mathcal{L}_{\text{total}} = \lambda_{\text{GT}} \cdot \mathcal{L}_{\text{GT}} + \lambda_{\text{PM}} \cdot \mathcal{L}_{\text{PM}} + \lambda_{\text{HM}} \cdot \mathcal{L}_{\text{HM}} + \lambda_{\text{AM}} \cdot \mathcal{L}_{\text{AM}}
\end{equation}

We identify the best weighing coefficients for the individual loss components through hyperparameter tuning on a held-out validation dataset.
We initialize the student model with the parameters from the best teacher model, as it has been shown in the distillation literature that it improves model performance and convergence steps compared to random initialization~\cite{wang2023distill}.
The teacher model is frozen during training, and only the student model is optimized using AdamW with linear learning rate warmup.
Additionally, early stopping based on validation loss is employed to prevent overfitting.

\descr{Student model results.}
The performance of the student model following the distillation procedure is summarized in Table~\ref{tab:eval_teacherserp}.
We find that the student model is able to approximate the performance of the teacher model reasonably well across most scam categories, indicating effective knowledge transfer through the distillation process.
Moreover, the student consistently outperforms both the data-driven baseline model (DistilBERT), which is trained solely on ground truth supervision without access to privileged information and the best-performing attributes or entities.
This improvement highlights the utility of incorporating LUPI from a more informative teacher, even when the student model operates under input constraints during inference time.

Notably, in certain categories such as C2 and C4, the student model achieves parity with the teacher model, indicating near-perfect distillation.
Conversely, the benefits of distillation are less pronounced in some categories, such as C1, where there is a notable performance gap between the student and teacher.
This gap can be attributed to the limitations of feature-based distillation under out-of-distribution (OOD) conditions, where performance deteriorates when the student encounters distributional shifts not seen during training.
The cross-validation strategy described in Section~\ref{sec:5foldcv} is intentionally designed to mimic an OOD evaluation scenario, where there is no category (domain) overlap between the training and testing splits, amplifying the effect of domain shift and approximating a worst-case evaluation scenario requiring hard generalization.

\descr{Takeaways.} In this section, we first empirically validated our intuition that SERP based representations offer valuable contextual signals for the task of toxicity prediction by training a privileged teacher model that leverages both the query and its associated SERP content, demonstrating significant performance gains over standard BERT-based regression baselines.
Next, we demonstrated that this privileged knowledge can be effectively transferred to a lightweight student model via feature distillation. 
The distilled student model, which relies solely on the query text, achieves consistently higher toxicity estimation and subsequently increased expansion factor across all five scam categories compared to the standard baseline.
Given our evaluation setup, which is deliberately designed to be challenging by excluding the scam category altogether from training in each fold, we argue that \systemname is well positioned to generalize to both known and emerging scam categories.
This generalization is crucial, as the primary utility of this keyword scoring model lies in its ability to surface queries indicative of novel or evolving scam categories and business verticals over time.
Additionally, the final model is based on a compact DistilBERT architecture, enabling real-time deployment for query ranking.

In the next section, we illustrate the real-world utility of our system by applying it to identify scam websites in the wild.
We further demonstrate the system’s capacity to generalize by discovering scam websites in previously unseen scam categories, thereby affirming its effectiveness in detecting emerging threats.

\section{Applying \systemname in the wild}
\label{sec:section_wild}
We previously validated the efficacy of \systemname by comparing it against static heuristics and data-driven baselines for sampling the most toxic queries on ground truth data.
Now, we use these toxic queries sampled by \systemname to discover new scam websites in the wild.
To this end, we use \systemname to rank the toxicity of the keywords collected in Section~\ref{sec:dataset_keywords} corresponding to the scam websites sourced from 49 different categories curated in Section~\ref{sec:categories}.
We 
select the top 20 ranked keywords from each category, resulting in a total of 980 distinct queries.
Following the methodology used in Section~\ref{sec:validation_dataset}, we use these 980 queries to collect SERPs from Google, Baidu, Bing, and Naver, resulting in a total of 271,161 websites.
The collected websites are passed through the oracle classifier, which labels 52,493 (19.3\%) of them as scams.
We refer to this set of newly identified fraudulent websites as the ``discovered scams.''

We present a summary of the discovered scams corresponding to the different categories curated in Section~\ref{sec:categories} in Table~\ref{tab:wild_categories}.
As expected, the highest number of scams were identified in the Crypto / Money (C2) category with 8,900 websites, followed by 
Adults / Gambling (C3) category, totaling 6,822 websites.
Additionally, \systemname helped us identify scams in underexplored categories, i.e., Education (C6) and Business / Admin (C5).
This demonstrates the generalizability of our system beyond traditionally well-studied scam verticals such as Shopping, cryptocurrency and investment scams~\cite{li2023double}. 
\begin{table}[]
  \centering
\begin{tabular}{|l|l|}
\hline
\textbf{Scam Type}      & \textbf{Identified Scams} \\ \hline
Shopping / Fashion (\textbf{C1}) & 6,034                     \\ \hline
Crypto / Money (\textbf{C2})     & 8,900                     \\ \hline
Adults / Gambling (\textbf{C3})  & 6,822                     \\ \hline
Medical / Pharmacy (\textbf{C4}) & 1,932                     \\ \hline
Pets / Animals (\textbf{C5})    & 2,737                    \\ \hline
Electronics (\textbf{C6})        & 3,249                     \\ \hline
Business / Admin (\textbf{C7})   & 3,459                     \\ \hline
Education (\textbf{C8})          & 3,841                     \\ \hline
Marketing / Sales (\textbf{C9})  & 2,305                     \\ \hline
Online Marketplace (\textbf{C10}) & 2,756                     \\ \hline
\end{tabular}
\caption{New scam websites identified by \systemname for each category.}
\label{tab:wild_categories}
\end{table}

\descr{User Impact Evaluation.}
The primary goal of \systemname is to build a proactive pipeline that can detect new scam websites, so that security systems can alert users when they access the scam website, which can happen through gateways other than search engine queries themselves (e.g., social media ads, spam emails, etc).
We utilize scam websites occurring on SERP results primarily as an index to discover these websites.
However, we find that based on different search engines, these scam websites do end up appearing in the top search results. 13.98\% of scam websites appear in the top 20 Google Search results, whereas the ratio is much higher on Bing (22.22\%), Naver (29.1\%), and Baidu (45.3\%).

While our pipeline incorporates a robust oracle for scam identification, dedicated downstream systems such as Scamdog Millionaire~\cite{kotzias2023scamdog} and ScamNet~\cite{bitaab2025scamnet} can serve as alternative mechanisms for classifying the websites discovered by \systemname, when available.
The stream of websites discovered by \systemname has a higher likelihood of being scams, and thus can be prioritized by downstream detection systems over randomly sampled websites and websites sourced through reactive sampling approaches.
Since there is no ground truth attached to the new scam websites discovered by \systemname, we use third-party scores such as ScamAdviser, Trustpilot, and social media profiles of the websites for further verification.%
A detailed analysis of each of the validation steps follows.

\subsection{ScamAdviser Validation}
\begin{figure}
\centering
\includegraphics[width=0.8\columnwidth]{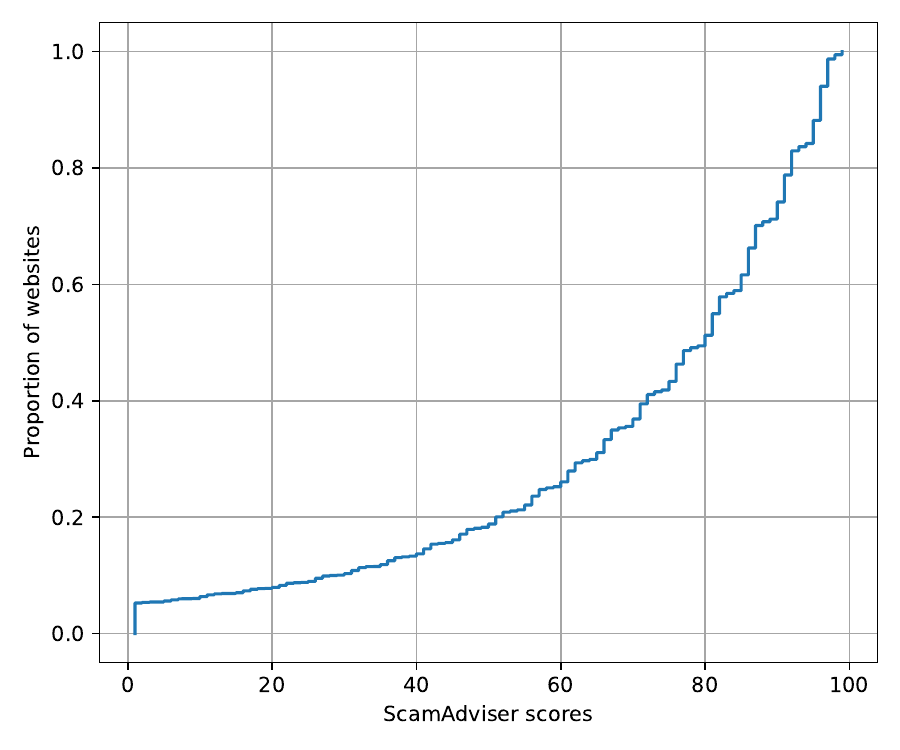}
\caption{ScamAdviser score of the scam websites identified by \systemname.}
\label{fig:validation_scamadviser}
\end{figure}

Prior work~\cite{kotzias2023scamdog} used scores from third-party security services such as ScamAdviser as a filtering mechanism and validation mechanism for building ground truth datasets of scam websites and benign websites.
We specifically borrow the ScamAdviser threshold used by the work, i.e., 85, for selecting benign websites as part of their ground truth for our analysis.
We query the ScamAdviser API with the discovered scams and record their scores (ranging from 0 to 100).
Figure~\ref{fig:validation_scamadviser} shows the cumulative distribution of ScamAdviser scores for the scam websites identified by \systemname.
The CDF reveals that approximately 62\% of the scam websites receive a score below the benign threshold (i.e., 85), and alarmingly 25\% of the websites receive the labeling of \textbf{\textit{Very Likely Unsafe}}, \textbf{\textit{Likely Unsafe}}, and \textbf{\textit{Caution Recommended}} by the ScamAdviser system (i.e., score below 60).

\subsection{TrustPilot Validation}
Similarly, prior work~\cite{kotzias2023scamdog} has used TrustPilot ratings as a mechanism for filtering and identifying scam websites.
We collect the TrustPilot profiles of the identified scam websites, collecting 5,540 TrustPilot ratings (i.e., 10.55\% of the websites).
This is in line with findings inthe  literature where almost 90\% of scam websites are not listed in Trustpilot as business profiles.
Additionally, 52 of these websites were reported by TrustPilot to be breaching their community guidelines, which is indicative of the websites using fake reviews or building inauthentic indicators of trust.
3,366 scam websites (60.75\%) receive a Trustscore rating of \textbf{\textit{Poor}}, indicating that most of the TrustPilot reviews received by scam websites are negative.
Only 665 (12.03\%) of the scam websites receive a Trustscore rating of ``Excellent,'' which has been used by prior works as a filtering criteria for benign websites~\cite{kotzias2023scamdog}.
Although some of the higher TrustPilot scores can be attributed to false positives from the oracle, others could reflect illegitimate reviews that bypassed Trustpilot’s detection system and therefore warrant further analysis using advanced Natural Language Processing (NLP) based methods.

\subsection{Instagram Followers Validation}
\begin{figure}
\centering
\includegraphics[width=0.8\columnwidth]{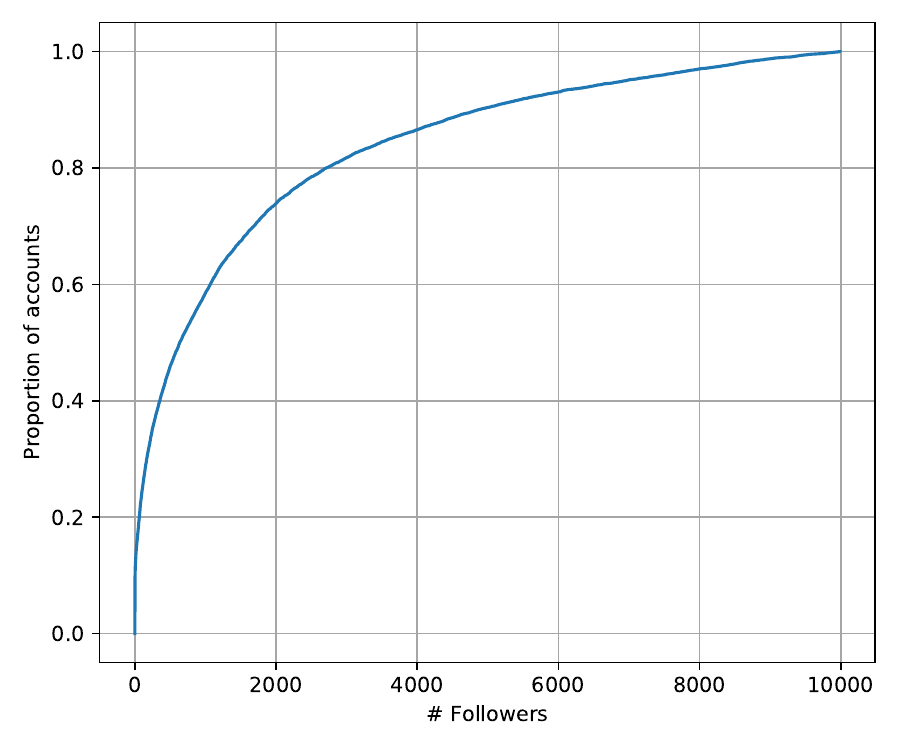}
\caption{Instagram follower counts of the scam websites identified by \systemname.}
\label{fig:validation_igfollower}
\end{figure}

Another important social signal associated with both genuine and scam websites is their social media presence on Facebook and Instagram.
Prior work~\cite{bitaab2023beyond} identified a lack of social media presence, relatively lower social media account age, and a lower number of followers as strong indicators of scam websites.
We leverage these findings to understand the composition of social media followers of the discovered scam websites.
Of the 52,493 discovered scam websites, only 12,866 (24.5\%) of the websites have a linked Instagram profile.
Of these, 1,082 (8\%) of the linked Instagram profiles are broken (provided Instagram links do not resolve to a valid profile), corroborating the findings in prior work~\cite{bitaab2023beyond}.
We plot the cumulative distribution function (CDF) of the Instagram follower counts associated with the scam websites in Figure~\ref{fig:validation_igfollower}.
The distribution reveals a strong right skew, over 90\% of scam-linked Instagram accounts have fewer than 5,000 followers, and approximately 50\% fall below the 500 follower mark.
Notably, 25\% of these accounts have fewer than 100 followers, which might be indicative of relatively newer scam operations.
Therefore, \systemname can be effective in identifying scam websites proactively before they expand their operations and incur significant financial damages.

\descr{Takeaways.}
\systemname enabled large-scale discovery of scam websites across a diverse set of categories, including those that have been traditionally underrepresented in prior work.
Moreover, the discovered websites offer high-yield websites for downstream detection systems, enabling more targeted and efficient classification than reactive or random sampling approaches.
While the external validation mechanisms (e.g., ScamAdviser and Trustpilot) provide useful corroboration of scam likelihood, they are inherently limited by outdated scoring heuristics and incomplete coverage of the underlying system.
These signals often lag behind emerging threats and can be subject to manipulation, meaning they do not fully capture the true prevalence of scams surfaced by \systemname. As such, our validation results likely represent a conservative estimate of the system’s effectiveness.

\section{Related Work}
In this section, we review relevant work on approaches to sample malicious websites in relevant security problems, and existing systems to classify scam websites.

\descr{Keyword-based approaches to sampling malicious websites.}
Prior efforts for discovering scam websites predominantly rely on domain-specific keyword lists enriched through static methods.
These keyword lists are usually constructed using brittle heuristics or manual curation, limiting their ability to generalize across diverse scam categories or adapt to new scam campaigns.
Srinivasan et al.~\cite{srinivasan2018exposing} derive search keywords from an initial corpus of TSS domains via TF‑IDF weighting and subsequently perform a post-hoc analysis that correlates query‑level traffic characteristics to the toxicity of the TSS sites discovered.
Similarly, Liu et al.~\cite{liu2023understanding} employ a topic modeling approach~\cite{blei2003latent} on the HTML content of known scam websites to extract representative search queries, which are then used to identify a diverse set of TSS websites~\cite{liu2023understanding,srinivasan2018exposing}.
Another line of work focuses on the detection of Fake Removal Advertisement (FRAD) websites by crafting a template of removal-intent queries (e.g., \textit{how to remove} \textbf{virus}) to surface malicious domains targeting users searching for malware remediation~\cite{koide2020never}.
EvilSeed~\cite{invernizzi2012evilseed} employs TF-IDF-based sampling on the output of Yahoo's Term Extraction API to identify popular topics and terms misused by miscreants to drive search engine traffic.
Leontiadis et al.~\cite{leontiadis2011measuring} sample illicit online drug selling websites by iteratively expanding a manually curated seed set of 218 drug-related queries to identify search-redirection attacks in the illicit online pharmacy ecosystem.
A common theme across all of these works is that they focus on a single type of malicious behavior (e.g., TSS, FRAD etc.), while our work focuses on studying a diverse set of scam categories (10), thus providing a more general and comprehensive understanding of guided search-based discovery of scam websites.
Additionally, we provide a framework for systematic baselines and a rigorous evaluation methodology for assessing the efficacy of different keyword-based sampling strategies.

\descr{Scam detection systems.}
Recent work in scam detection has focused on developing scalable and effective systems for identifying fraudulent e-commerce websites across a wide range of online services.
Bitaab et al.\cite{bitaab2023beyond} present the first-ever detection system that goes beyond the scope of traditional phishing websites to detect fraudulent e-commerce sites by leveraging a comprehensive infrastructure including crawling, feature extraction, and supervised learning.
Kotzias et al.\cite{kotzias2023scamdog} introduce Scamdog Millionaire, a detection system designed for identifying fraudulent ``e-shops'' at scale for a commercial security system.
Towards developing explainable scam-detection systems, ScamNet\cite{bitaab2025scamnet} incorporates the reasoning capabilities of Large Language Models (LLMs) for explainable scam detection, producing human-understandable reasoning behind classification decisions.
ScamMagnifier\cite{Bitaab2025SCAMMAGNIFIERPT} leverages the shared payment infrastructure of organized scam operations, offering insights into coordinated scam ecosystems and enabling broader detection strategies. 
Collectively, these works demonstrate the evolution of scam detection systems from static classifiers to explainable, campaign-aware frameworks prioritizing different phases and \textit{modus operandi} of fraudulent websites.

\section{Discussion and Conclusion}
We presented \systemname, a novel data-driven framework for mining high-toxicity search queries in order to maximize the yield in discovery of fraudulent e-commerce websites via Search Engine Results Pages (SERPs).
Through a comprehensive evaluation of existing search traffic features and query understanding heuristics, we demonstrated that prior approaches perform inconsistently across different scam categories and yield suboptimal sampling of toxic queries.
To overcome these limitations, we developed a category-agnostic, data-driven keyword scoring model powered by real-world SERP feedback.
We rigorously validate the efficacy of \systemname on ranking the most toxic set of keywords from a pool of 1.5 million candidate keywords, demonstrating that the search keywords identified by \systemname outperform both supervised and unsupervised baselines in the task of discovering new scams.
We further applied \systemname in the wild to discover 52,493 previously unreported scam websites spanning ten distinct scam categories, thereby demonstrating its practical effectiveness in real‐world settings.

Our paper takes a first step towards developing systems to proactively identify scam websites by leveraging data-driven query mining.
We believe that the end-to-end system pipeline proposed in our approach serves as a new paradigm for security analysts to stay ahead of scam adversaries in the rapidly evolving scam ecosystem.
By shifting from reactive and manually crafted keyword lists to proactive, SERP-informed query mining, \systemname presents itself as a sound framework for longitudinal discovery of scam websites.
We believe that \systemname (which we make publicly available\footnote{https://doi.org/10.5281/zenodo.17049965}) will be an effective tool for enabling security analysts to proactively identify scam websites.

\descr{Ethics.}
All datasets used in this work were either publicly released by other researchers or reported by users on public scam detection forums and communities.
Additional sources of data (i.e., Google Keyword Suggestions and the SERP results) are collected from the official API providers, following those API's terms of service.
This work is not considered human subjects research by our institution, since we do not interact with humans and do not collect any private information.
Although we issue potentially toxic queries to the search engines to measure toxicity and identify new scam websites during experimentation, all requests are subject to strict rate limits and comply with the providers acceptable-use policies to ensure responsible use of the crawling infrastructure.
While \systemname is developed to aid the proactive detection efforts, the query ranking outputs can be misused by adversaries as part of curating their own ``content blacklist'' to avoid using these keywords in their page source, in an effort to prevent these webpages from being indexed by the search engines within the context of the identified toxic keywords. 
However, eliminating such restrictions on the content published in the scam websites compromises the intended tactics of the scam operations and ultimately decreases the effectiveness and presentation of the scam websites.%

\descr{Design implications.}
\systemname's design is motivated with a focus towards practical adoption in real-world security systems.
The core component, a light-weight regression model designed on DistilBERT has a very small memory footprint (268 Megabytes), and can be served with inference on a standard device with minimal GPU configurations or even by offloading the model weights to a CPU.
Input to \systemname is a list of queries related to a scam website or a brand vertical, which can be sourced from anywhere.
During the design of our system, these keywords are sourced from Google Ads keyword Suggestions to approximate real-world traffic and issue queries resembling real-world search behavior, but \systemname can be fed any input stream of keywords (including results from other keyword extraction systems). %
Other components of our work can be easily integrated into existing security subsystems (e.g., alternate and dedicated oracles to classify the new websites discovered via SERP can be used.)
Moreover, the end-to-end system pipeline, as illustrated in Figure~\ref{fig:system_pipeline} can be used to enable continuous discovery of new scam websites with minimal human supervision.
Starting from a small seed of scam websites to query identification and ranking, the newly identified scam websites from \systemname can be fed back to the end-to-end system to identify websites with evolved scam tactics, in a loop.

\descr{Extending the capabilities of \systemname.}
\systemname is designed to be campaign-agnostic as we demonstrated that it generalizes well across the scam categories identified in the paper, discovering new scam websites across 10 categories (in Table~\ref{tab:eval_teacherserp}).
To adapt Loki to any new type of scam campaigns or categories, we first need to identify a few websites (either scams or benign) that are related to an e-commerce campaign or business category. 
These websites can then be passed through our pipeline in Figure~\ref{fig:system_pipeline}: first, Google Ad Keywords for getting keyword suggestions, and the student Loki model to rank the targeted set of keywords to be issued for SERPs.
Finally, the ranked set of keywords can be queried via SERP APIs or search engine interfaces and passed through an oracle to identify potential scam websites. 
This way, starting from a seed set of websites related to a category or campaign, we can easily identify new scam websites related to the campaign.

\descr{Limitations.}
While our system demonstrates strong generalizability and improved discovery of scam websites across diverse categories, it is not without limitations.
First, the discovery enabled by our approach is inherently constrained to websites that are indexed by the search engines studied.
Scam websites that rely on alternative discovery mechanisms (e.g., direct social media links or messaging platforms) or intentionally exclude themselves from search engine indexing remain beyond the reach of \systemname.
Finally, scammers increasingly employ social engineering, multimedia-driven lures, or multilingual and multi-modal deception strategies that may not rely on conventional linguistic constructs captured by search keywords.
\systemname's purely keyword-based keyword discovery will miss such scams, limiting the system’s applicability in adversarially evolving threat landscapes.

\descr{Acknowledgments.} We would like to thank the anonymous reviewers for their feedback and Jason Redick for his help on data collection efforts.
This work was supported by the National Science Foundation under Grants CNS-2419829, CNS-2346845, CNS-2247868, and CNS-1942610.

\begin{small}
\bibliographystyle{abbrv}
%\bibliography{bibliography}

\end{small}
\clearpage
\begin{appendices}

\section{Data collection: Scam websites}
\label{appendix:data_scam}
A brief description of each of the sources for scam websites follows:
\begin{itemize}
    \item \textbf{BeyondPhish}: BeyondPhish represents the first large-scale effort to systematically collect and analyze fraudulent e-commerce websites~\cite{bitaab2023beyond}. 
    The dataset is constructed by leveraging user-submitted reports from the popular subreddit \textit{/r/scams}, which are subsequently verified by security experts who manually inspect and label live websites as scams.
    The study's definition of Fraudulent e-Commerce Websites (FCWs) is well-aligned with our operational definition of scam websites.
    Scam websites present in the dataset expand over 6 different categories, such as shipping scams, adult services, etc.
    This publicly available dataset (released in 2023) contains 4,123 scam websites, of which only 1,315 websites were active during the time of our evaluation.
    Although the HTML source code is available for all entries, our pipeline depends on features extracted from live websites; hence, we restrict our usage to the subset of currently active domains.

    \item \textbf{The Scam Directory:} The Scam Directory is a community-led effort to label scams and alert users for online protection.
    The platform maintains a catalog of different categories of scams, e.g, fashion scams, non-delivery scams, sport scams, etc.
    Recent work on classifying scam websites~\cite{kotzias2023scamdog} incorporates this dataset as part of building their e-commerce scam classifier model.

    \item \textbf{ScamGuard:} ScamGuard is a public platform that allows consumers to report potential scam websites and publish their list of identified scam domains.
    This platform covers a broad spectrum of 8 different scam categories, e.g., investment scams, fake stores, crypto scams, pet scams, etc.
    Similar to The Scam Directory, prior work~\cite{kotzias2023scamdog} utilizes this dataset as part of curating their labeled dataset for training a Machine Learning (ML) based scam classifier.
\end{itemize}

\section{Branded Keyword classifier}
\label{appendix:branded_classifier}
Classifying keywords as branded or un-branded is a well-recognized task within the Search Engine Optimization (SEO) industry, where it is used to inform marketing strategies and search traffic analysis.
However, despite its practical significance, there exists no publicly available dataset, benchmark, or standardized model for this classification task.
Existing solutions are predominantly proprietary and lack transparency, thereby limiting their applicability in academic or open source contexts.
To address this gap, we develop a custom classification pipeline to identify and filter branded keywords as part of our data preprocessing stage.

We begin by constructing a validation dataset to guide the classification of branded versus un-branded keywords. 
To this end, we randomly sample 4,000 query keywords from the Google Ads Keyword Suggestions dataset (described in Section~\ref{sec:dataset_keywords}), covering 25 unique domains.
For each domain, we manually label both branded and un-branded keywords, resulting in a dataset comprising 2,064 branded keywords and 1,936 un-branded keywords.

Next, we leverage the zero-shot classification capabilities of Large Language Models (LLMs) to classify if the keywords are branded.
The rationale behind this approach is that foundation models, pretrained on extensive and diverse internet corpora, inherently capture a wide range of factual knowledge, including the names of commercial brands, products, and organizations.
With carefully designed prompts and a clear definition of ``brandedness,'' we expect LLMs to be effective at identifying branded keywords without requiring task-specific fine-tuning.
Recent security works~\cite{liu2024less} leverage LLMs' internal representations to associate brand and domain information for the purpose of reference-based phishing detection without a pre-defined reference list.

This classification task is performed in a zero-shot manner, relying solely on prompt engineering rather than supervised training.
Recent work has demonstrated the utility of LLMs for related classification problems, including zero-shot detection of scam websites~\cite{bitaab2025scamnet} and few-shot classification of e-commerce websites~\cite{tsai2025harmful}.
Although the task is unsupervised, we use our manually annotated validation dataset to evaluate different prompt templates (i.e., verbalizers), and select the one that yields the highest classification accuracy for branded keyword identification.
We implement the FLAN-T5-XXL model for our task, a state-of-the-art instruction-tuned language model which has shown state-of-the-art in multiple text classification benchmarks\cite{google_flan,chung2022scaling}, and novel tasks such as claim-based stance detection~\cite{paudel2024enabling}.
We experiment with different prompts for the task definition, and find that the following task definition in Figure~\ref{fig:prompt_setup} performs the best on the validation dataset.
Using this configuration, the FLAN-T5-XXL model achieves an F1 score of 90\%, which we consider sufficiently accurate for the purposes of filtering branded keywords within our pipeline.

\begin{figure}
\begin{mdframed}[style=MyFrame,nobreak=true]
\begin{quote}
Classify if the provided user query is a branded query, or un-branded query.
Branded Queries are ones that includes a specific company or brand name, indicating the user is already familiar with the brand. (e.g. nike shoes, apple watch etc).
Un-branded Queries are ones that does not mention any specific brand, suggesting the user is looking for general information without a particular company in mind (e.g. shoes for mens, wrist watch etc).

Remember that mention of brands can happen at any position within the queries.
Additionally, some brand name can be entirely generic terms, in which case use of context is necessary for proper classification.

Return your response as 1 for branded and 0 for un-branded.
User query: {0}

Response:
\end{quote}
\end{mdframed}
\caption{Prompting structure for branded keyword filtering.}
\label{fig:prompt_setup}
\end{figure}

\section{Oracle Features}
In this section, we list the features (previously discussed in Section~\ref{sec:oracle_model}) used in the development and validation of the oracle model for classifying whether a website is scam or benign.
\begin{table}[ht]
\centering
\small
\begin{tabular}{|l|l|l|}
\hline
\textbf{Category} & \textbf{Feature} & \textbf{Type} \\ \hline

\multirow{6}{*}{Domain Ranking}
 & majestic\_refips        & N \\
 & majestic\_refsubnets    & N \\
 & majestic\_tldrank       & N \\
 & tranco                  & N \\
 & majestic                & N \\
 & cisco                   & N \\ \hline

\multirow{19}{*}{DNS Features}
 & dns\_has\_mx                         & B \\
 & dns\_num\_mx                         & N \\
 & dns\_has\_cname                      & B \\
 & dns\_num\_cname                      & N \\
 & dns\_has\_dname                      & B \\
 & dns\_num\_dname                      & N \\
 & dns\_has\_hinfo                      & B \\
 & dns\_num\_hinfo                      & N \\
 & dns\_has\_aaaa                       & B \\
 & dns\_num\_aaaa                       & N \\
 & dns\_has\_ns                         & B \\
 & dns\_num\_ns                         & N \\
 & dns\_has\_rp                         & B \\
 & dns\_num\_rp                         & N \\
 & dns\_has\_soa                        & B \\
 & dns\_num\_soa                        & N \\
 & dns\_has\_txt                        & B \\
 & dns\_num\_txt                        & N \\
 & dns\_domain\_verification\_count     & N \\ \hline

\multirow{6}{*}{URL Based}
 & tld             & C \\
 & cheap\_tld               & B \\
 & domain\_subwords         & N \\
 & url\_has\_hyphen         & B \\
 & url\_has\_digit          & B \\
 & url\_subdomain\_count    & N \\ \hline

\multirow{8}{*}{WHOIS Based}
 & domain\_age            & N \\
 & time\_to\_expiry       & N \\
 & registrar\_name        & C \\
 & is\_cheap\_registrar   & B \\
 & registrar\_country     & C \\
 & registrant\_country    & C \\
 & privacy\_protected     & B \\
 & free\_email\_provider  & B \\ \hline

\multirow{29}{*}{Content Based}
 & facebook\_profile\_linked         & B \\
 & twitter\_profile\_linked          & B \\
 & instagram\_profile\_linked        & B \\
 & youtube\_profile\_linked          & B \\
 & pinterest\_profile\_linked        & B \\
 & tiktok\_profile\_linked           & B \\
 & presence\_of\_contact\_link       & B \\
 & num\_mailto\_links                & N \\
 & num\_telephone\_links             & N \\
 & num\_whatsapp\_links              & N \\
 & review\_system\_linked            & B \\
 & has\_app\_store                   & B \\
 & has\_review\_widget               & B \\
 & num\_links                        & N \\
 & num\_internal\_links              & N \\
 & num\_external\_links              & N \\
 & num\_img\_tags                    & N \\
 & num\_iframe\_tags                 & N \\
 & num\_external\_http\_links        & N \\
 & num\_links\_with\_ip              & N \\
 & presence\_work\_with\_us\_link    & B \\
 & presence\_cookie\_consent\_notice & B \\
 & trustpilot\_present               & B \\ \hline

\end{tabular}
\caption{Feature Categories and Their Types (B = Boolean, N = Numeric, C = Categorical).}
\label{tab:feature_categories}
\end{table}
\end{appendices}

\end{document}